\documentclass[12pt,letterpaper]{article}
             
\usepackage{jheppub} 
                     
\usepackage{color}
\usepackage{setspace}
\usepackage{slashed}
\usepackage{makecell}
\usepackage{multirow}
\usepackage{wasysym}
\usepackage{physics}
\usepackage{changepage}
\usepackage{arydshln}     
\usepackage{scalerel} 
\usepackage{slashed}
\usepackage{comment}
\usepackage{amsmath,amsthm,amssymb,amsfonts,graphicx,hyperref,ccaption,amstext}
\usepackage{enumitem}
\usepackage[makeroom]{cancel}

\usepackage{xcolor}
\usepackage{lscape}

\def\beq{\begin{equation}}
\def\eeq{\end{equation}}
\def\bea{\begin{eqnarray}}
\def\eea{\end{eqnarray}}

\newcommand{\la}{\langle}
\newcommand{\ra}{\rangle}

\title{\boldmath Conformal Freeze-In,  Composite Dark Photon,  and Asymmetric Reheating}

\author[a,b]{Wen Han Chiu,}
\author[a,b,c]{Sungwoo Hong,}
\author[a,b,d]{Lian-Tao Wang}

\affiliation[a]{Department of Physics, University of Chicago, Chicago, Illinois 60637, U.S.A.}
\affiliation[b]{Enrico Fermi Institute, University of Chicago, Chicago, Illinois 60637, U.S.A}
\affiliation[c]{Argonne National Laboratory, Lemont, IL 60439, USA}
\affiliation[d]{Kavli Institute for Cosmological Physics, University of Chicago, Chicago, Illinois 60637, U.S.A.}
\emailAdd{wenhan@uchicago.edu}
\emailAdd{sungwooh@uchicago.edu}
\emailAdd{liantaow@uchicago.edu}
\abstract{Large classes of dark sector models feature mass scales and couplings very different from the ones we observe in the Standard Model (SM). Moreover, in  the freeze-in mechanism, often employed by the dark sector models, it is also required that the dark sector cannot be populated during the reheating process like the SM. This is the so called asymmetric reheating.
Such disparities in sizes and scales often call for dynamical explanations. In this paper, we explore a scenario in which slow evolving conformal field theories (CFTs) offer such an explanation. Building on the recent work on conformal freeze-in (COFI), we focus on a coupling between the Standard Model Hypercharge gauge boson and an anti-symmetric tensor operator in the dark CFT. We present a scenario which dynamically realizes the asymmetric reheating and COFI production. With a detailed study of dark matter production, and taking into account limits on the dark matter (DM) self-interaction, warm DM bound, and constraints from the stellar evolution, we demonstrate that the correct relic abundance can be obtained with reasonable choices of parameters. The model predicts the existence of a dark photon as an emergent composite particle, with a small kinetic mixing also determined by the CFT dynamics, which correlates it with the generation of the mass scale of the dark sector. At the same time, COFI production of dark matter is very different from those freeze-in mediated by the dark photon. This is an example of the physics in which a realistic dark sector model can often be much richer and with unexpected features.}

\begin{document}
\maketitle
\flushbottom

\section{Introduction}

Dark sector models (see \cite{Battaglieri:2017aum,Alexander:2016aln} for overviews and relevant references) offer promising avenues beyond the weakly interacting massive particle (WIMP) paradigm. The mass scales in such models are often much lower than those we have in the Standard Model (SM). For phenomenological reasons, their coupling to the SM will need to be strongly suppressed as well. The production of dark matter is usually very different from the freeze-out mechanism commonly employed by the WIMP. Instead, a freeze-in mechanism \cite{McDonald:2001vt, Hall:2009bx} is often invoked. The small coupling between the dark sector and the SM ensures they are not in thermal equilibrium. At the same time, the dark sector can't be populated during the reheating process like the SM. Implementing such an asymmetric reheating is a requirement for the success of a freeze-in model. While a simple parameterization with low energy degrees of freedoms is usually enough for phenomenological studies, such an array of different scales and small parameters usually call for dynamical explanations. 

It is well known that large scale separation is present in theories which are nearly scale invariant, that is those close to being a conformal field theory (CFT). Starting at some UV scale where the theory is approximately conformal, a small deformation can lead to the emergence of an infrared scale which is exponentially lower than the UV scale. Hence, such CFTs are natural candidates for dark sector models. The deformation are generically present, for example, through the coupling with the SM. The small couplings required in such scenarios can be generated from scale separation as well. Motivated by this, there have been recent works \cite{Hong:2019nwd, Hong:2022gzo} studying the conformal freeze-in  (COFI) process where the dark sector is conformal.  The deformation would eventually lead to the confinement of the dark CFT, generating a mass gap, $m_{\rm gap}$. A natural candidate of dark matter is one of the low lying composite resonances. Making an analogy with quantum chromodynamics (QCD), we will consider a dark matter candidate which is similar to the pion, with mass about one or two orders of magnitude below $m_{\rm gap}$. 

Building on the set of work on COFI, we set out to build a complete model which leads to the production of dark matter with the correct relic abundance. We consider a coupling (a portal) between the SM hypercharge gauge boson and an antisymmetric tensor operator in the dark sector CFT, which is the main driver for the COFI dark matter production. Other connections with the dark sector could also be (and have been \cite{Hong:2022gzo}) considered. We offer a dynamical explanation of the smallness of the coupling between the SM and the dark CFT sector. In addition, we propose a scenario in which asymmetric reheating can be realized. Dark sector models are also subject to a host of astrophysical and cosmological constraints, including DM self-interaction, warm DM bound, and star cooling bounds. Taking these into account, we identify models in which correct dark matter relic abundance can be generated. 

Our model predicts the existence of the dark photon as a composite vector meson in the dark sector with mass close to $m_{\rm gap}$.  The portal coupling introduced earlier will transform into a kinetic mixing between the dark photon and the SM hypercharge gauge boson in the IR once the conformal dark sector confines. The smallness of this coupling is explained by a large scale separation induced by a slow renormalization group (RG) running between $\Lambda$ and $m_{\rm gap}$. There is one important difference between our model and models with an elementary dark photon. While the freeze-in is mediated by the elementary dark photon in the latter case, the dark photon does not play a role during the COFI production. Hence, the relation between the relic abundance and the mass and coupling of the dark photon is very different, as illustrated in \autoref{Fig:relic_density_constraints_IR_dom}.

The rest of the paper is structured as follows. In \autoref{sec:setup}, we describe our theory and its IR effective theory. In particular, in \autoref{subsec:UV theory}, we discuss UV theory and explain how the small coupling and asymmetric reheating required for the non-thermal freeze-in production can be achieved. Then, \autoref{subsec:IR EFT} is devoted to describing the IR effective theory of dark matter and composite dark photon and mass gap generation. In \autoref{sec:Pheno}, we present detailed analysis of dark matter phenomenology, including freeze-in production, cosmological evolution, and various observational constraints. We then conclude in \autoref{sec:conclusion}. Several technical details are relegated to appendices. Dynamical small mass scale generations in COFI theories are explained in \autoref{sec:dsScales}. 5d dual picture of 4d COFI theories via AdS/CFT correspondence is described in \autoref{app:AdS/CFT}. Production of the dark sector in its hadronic phase (as opposed to conformal phase) can occur when $T < m_{\rm gap}$ during the production and some details are presented in \autoref{sec:HadProduction}. Details of rate computations needed for COFI production are discussed in \autoref{app:COFI_computation}. Finally, useful ingredients of stellar evolution bounds for our theory are summarized in \autoref{app:star_cooling}.

\section{The Setup}
\label{sec:setup}

In this section, we introduce our theory and describe some of its key features. 
Our discussion in this section is mainly in the language of 4d QFT (CFT). Via the AdS/CFT correspondence, our theory admits a weakly coupled 5d gravity description which is presented in \autoref{app:AdS/CFT}. 
In addition, the production and evolution of the dark sector in cosmology and its phenomenology will be discussed in detail in \autoref{sec:Pheno}.

We are primarily interested in studying the conformal freeze-in production \cite{Hong:2019nwd, Hong:2022gzo} of the conformal dark sector coupled to the SM via a tensor interaction
\beq
\mathcal{L}_{\rm \scriptscriptstyle COFI} \supset \frac{\lambda}{\Lambda^{d-2}} B_{\mu\nu} \mathcal{O}^{\mu\nu}. \nonumber
\eeq
Here, $B_{\mu\nu}$ is the field strength of $U(1)_Y$ gauge boson in the SM, and we assume $\Lambda \sim \mathcal{O} (1)$ TeV and $\lambda \ll 1$ as is the norm for freeze-in. 
Readers interested in phenomenology of this theory may skip \autoref{subsec:UV theory} and jump directly to \autoref{subsec:IR EFT}. \autoref{subsec:UV theory} (and \autoref{app:AdS/CFT}) is devoted to the description of microscopic theory which, through a cascade confinement, addresses the question of asymmetric reheating and results in the above effective theory, the starting point of our phenomenological study in the rest of the paper.

\subsection{UV theory and asymmetric reheating}
\label{subsec:UV theory}

\begin{figure}[h]
  \centering
  \includegraphics[width=0.75\textwidth]{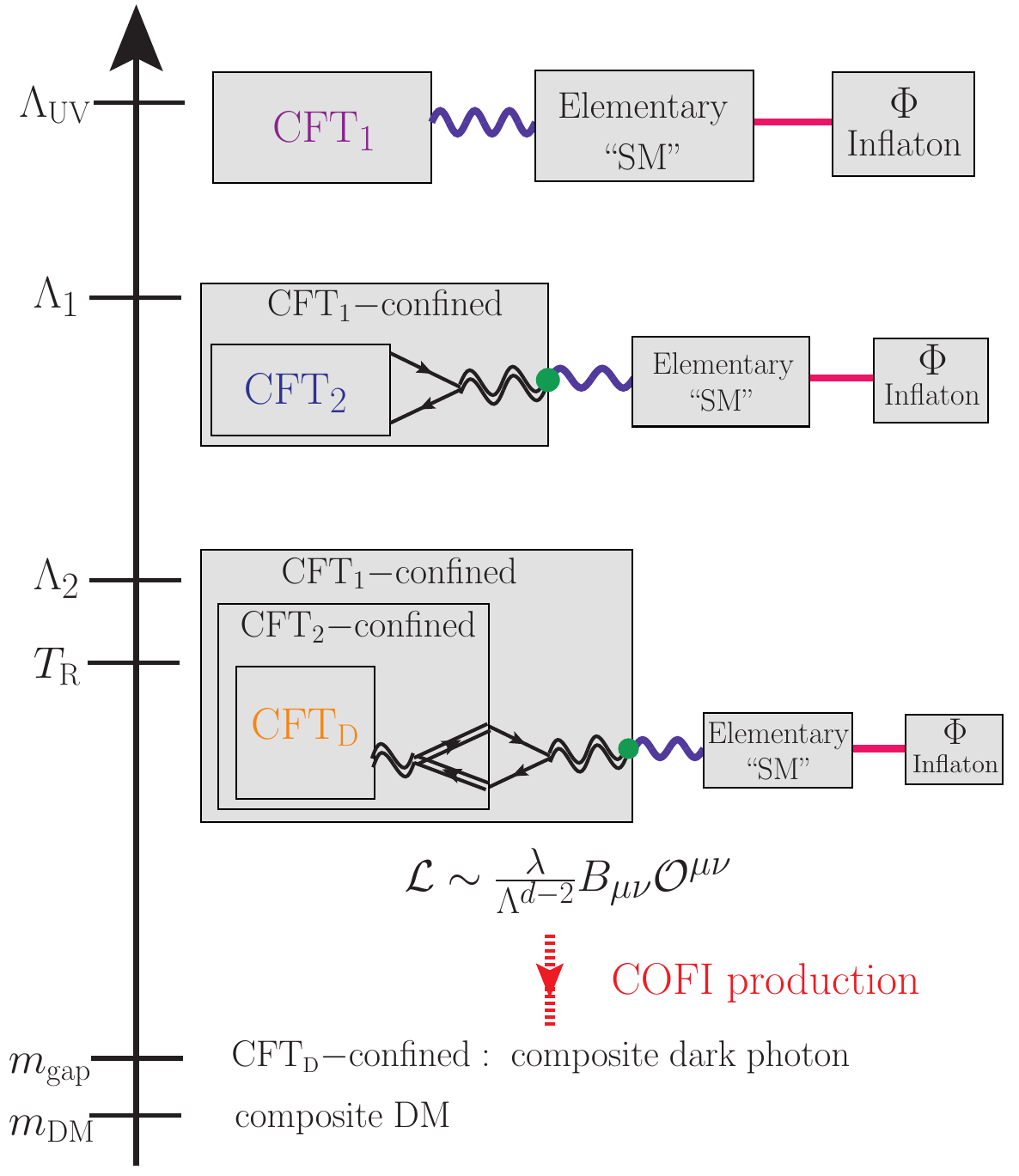}
  \caption{Our theoretical setup and its RG evolution. Cascade of confinement results in small effective coupling between the SM and dark CFT and asymmetric reheating.}
  \label{Fig:big piccture}
\end{figure}
In this section, we describe our UV theory and its RG evolution in a form of cascade confinement. The overall picture is depicted in \autoref{Fig:big piccture}.

In the UV, our theory consists of a sector of CFT (denoted as $\text{CFT}_1$) coupled to a sector of elementary (as opposed to composite) particles. The elementary sector includes the inflaton $\Phi$ and a copy of the SM particle contents. 
The relevant particle contents and their interactions can be summarized by
\beq
\mathcal{L}_{\rm \scriptscriptstyle UV} = \mathcal{L}_{\rm \scriptscriptstyle CFT_1} + \mathcal{L}_\Phi + \mathcal{L}_{\rm ext} \left( q, \ell, A_\mu \right) + \mathcal{L}_{\rm \scriptscriptstyle GW_1} + \mathcal{L}_{\rm \scriptscriptstyle RH} + \mathcal{L}_{\rm \scriptscriptstyle PC}
\label{eq:UV_theory}
\eeq
where
\begin{itemize}
\item[1.] $\mathcal{L}_{\rm ext}$ represents terms for the external  ``SM'' fields, $q = \text{quark}, \ell = \text{lepton}, A_\mu = \{ G_\mu^a, W_\mu^i, B_\mu \} = \text{gauge fields}$\footnote{Here, we do not include the SM Higgs as we wish to solve the EW hierarchy problem by treating the Higgs as composite. This, however, is not a necessary component of the model.}. These are not yet the SM fields. As described below, the SM fields are realized as admixtures of external (elementary) and composite states, i.e.~partial compositeness (PC), at energy scale below the confinement scale of $\text{CFT}_1$ by diagonalizing the elementary$\text{--}$composite mixing. Such a scheme is commonly used in the so called holographic Composite Higgs Model (CHM). For convenience, we will refer to the combination of external SM fields and the composite states they mix with as the CHM sector. 
\item[2.] $\mathcal{L}_{\rm \scriptscriptstyle GW_1} = \eta \mathcal{O}_{\rm \scriptscriptstyle GW_1}$ describes a scalar deformation responsible for the running of the $\text{CFT}_1$ and generation of stable mass gap in the IR. This is the CFT dual of the Goldberger-Wise stabilization mechanism in 5d \cite{Goldberger:1999uk} and more details can be found in \cite{Arkani-Hamed:2000ijo, Rattazzi:2000hs, Agashe:2016rle}.
\item[3.] $\mathcal{L}_{\rm \scriptscriptstyle RH}$ describes the interaction between the inflaton $\Phi$ and elementary fields, hence the reheating of the external sector. We emphasize that the inflaton is purely (or mostly) elementary with no (or little) composite mixture and hence it primarily couples only to the external sector. 
\item[4.] $\mathcal{L}_{\rm \scriptscriptstyle PC} = y_q \bar{q} \mathcal{O}_q + y_\ell \bar{\ell} \mathcal{O}_\ell + g A_\mu J_{\mu} + \frac{1}{M_{\rm pl}} h_{\mu\nu} T^{\mu\nu}_{\rm \scriptscriptstyle CFT_1}$ represents the linear interactions between the external fields and the CFT operators. When the $\text{CFT}_1$ confines in the IR, these will turn into the partial compositeness couplings between elementary fields and their composite partners.
\end{itemize}
The scalar deformation, $\mathcal{O}_{\rm \scriptscriptstyle GW_1}$, triggers RG running of $\text{CFT}_1$ and the conformal invariance breaking effect grows in the IR if it is a relevant operator. Eventually, at a scale $\Lambda_1$, it becomes an $\mathcal{O} (1)$ violation and leads to a spontaneous breaking of $\text{CFT}_1$ measured by the vacuum expectation value (vev) of $\mathcal{O}_{\rm \scriptscriptstyle GW_1}$. We assume that $\text{CFT}_1$ confines when this occurs. This event generates heavy composite particles which mixes linearly with the external fields. Upon diagonalizing this mass mixing, one gets mass eigenstates including massless states and these are identified as the SM particles. Heavy mass eigenstates correspond to the Kaluza-Klein (KK) excitations in the dual 5d picture.

In addition to the CHM sector described above, we assume that the confinement of $\text{CFT}_1$ also gives rise to a sector of composite ``preons'' which are singlets of the SM gauge group. These preons are similar to the quarks and gluons of QCD, and we assume that their dynamics bring them to an IR fixed point (denoted as $\text{CFT}_2$).~\footnote{Strictly speaking, the preon sector needs not be a CFT sector. For our purposes, it suffices that the dynamics of the composite preon sector has a slow RG running and an interacting IR fixed point at a much lower scale (this is our dark CFT). Provided this assumption, all our discussion below will be equally applicable.} The dynamics of $\text{CFT}_1$ and the phase transition may result in various couplings between the CHM sector and $\text{CFT}_2$. We assume that the dominant interaction is given by 
\beq
\mathcal{L}_{\scriptscriptstyle \Lambda_1} \supset \frac{1}{\Lambda_1} \rho_{B \mu\nu} \bar{\psi}_L \sigma^{\mu\nu} \chi_R + y_{\Psi} \bar{\psi}_L \mathcal{O}_R + y_{\chi} \bar{\mathcal{O}}_L \chi_R
\label{eq:Theory_at_Lambda_1}
\eeq
where $\rho_{B \mu}$ is the composite $U(1)_Y$ vector meson which couples to SM singlet composite fermions (preons) $\psi_L$ and $\chi_R$  via a dipole interaction. These latter SM singlet composite fermions couple to the $\text{CFT}_2$ through the linear mixing couplings. 
Since $B^{\mu}$ is external to the $\text{CFT}_1$, its coupling to the $\text{CFT}_2$ (which belongs to the composite sector) has to be through its mixing with composite partner $\rho_B^\mu$. This mixing is analogous to the $\gamma \text{--} \rho$ mixing realized in QCD and is given by $g/g_{1 s}$, where $g$ and $g_{1 s}$ are the $U(1)_Y$ gauge couping and composite coupling of confined $\text{CFT}_1$, respectively. See \cite{Agashe:2016rle} for more discussion.

Below $\Lambda_1$, the above theory will undergo RG flow and the details depend on the scaling dimensions of the fermionic operators of $\text{CFT}_2$, $\mathcal{O}_{L,R}$. Denoting the scaling dimensions of these as $d_L$ and $d_R$ respectively, we first consider $d_L, d_R > 5/2$.

\subsubsection*{(i) $d_L, d_R > 5/2$}

In this case, the linear couplings are irrelevant operators, and they decrease towards the IR. At some lower scale $\mu < \Lambda_1$, we get 
\beq
\mathcal{L}_{\scriptscriptstyle \mu < \Lambda_1} \supset \frac{1}{\Lambda_1} \rho_{B \mu\nu} \bar{\psi}_L \sigma^{\mu\nu} \chi_R + \tilde{y}_{\Psi} \left( \frac{\mu}{\Lambda_1} \right)^{d_R-\frac{5}{2}} \mu^{\frac{5}{2}-d_R} \bar{\psi}_L \mathcal{O}_R + \tilde{y}_{\chi} \left( \frac{\mu}{\Lambda_1} \right)^{d_L-5/2} \mu^{\frac{5}{2}-d_L} \bar{\mathcal{O}}_L \chi_R.
\label{eq:Theory_at_mu}
\eeq
We have defined a dimensionless coupling $\tilde{y}_\psi$ by $y_\psi = \tilde{y}_\psi \Lambda_1^{5/2-d_R}$, and similarly for $\tilde{y}_\chi$.

We imagine that at a scale $\Lambda_2 < \Lambda_1$, the composite $\text{CFT}_2$ confines, generating composite particles and yet another composite CFT denoted as $\text{CFT}_{\scriptscriptstyle \rm D}$. This $\text{CFT}_{\scriptscriptstyle \rm D}$ is the dark sector of our theory and carries dark $U(1)_{\scriptscriptstyle \rm D}$ global symmetry, hence reveals a coupling
\beq
\mathcal{L}_{\scriptscriptstyle \Lambda_2} \supset \mathcal{L}_{\scriptscriptstyle \text{CFT}_{\rm D}} + g_{\scriptscriptstyle \rm D} A_{{\scriptscriptstyle \rm D} \mu} J_{\scriptscriptstyle \rm D}^\mu + \cdots.
\eeq
Here, $A_{\scriptscriptstyle \rm D}^\mu$ is a composite vector meson of confined $\text{CFT}_2$ and simultaneously plays the role of external $U(1)_{\scriptscriptstyle \rm D}$ gauge field coupled to $\text{CFT}_{\scriptscriptstyle \rm D}$ current $J_{\scriptscriptstyle \rm D}^\mu$.
It also couples to a pair of composite fermions coming from $\mathcal{O}_{L,R}$ through a dipole interaction. 

We can use an interpolation relation between the fermionic CFT operators and canonically normalized composite fermion fields, $\mathcal{O}_R \sim \Lambda_2^{d_R-3/2} \sum_n c_n \psi_{\scriptscriptstyle \rm comp, R}^{(n)}$ and $\mathcal{O}_L \sim \Lambda_2^{d_L-3/2} \sum_n d_n \chi_{\scriptscriptstyle \rm comp, L}^{(n)}$ \footnote{The sum is over the tower of composite fermions. $c_n$ and $d_n$ denote the ``form factor''s.}, to obtain an effective action at $\Lambda_2$
\bea
\mathcal{L}_{\scriptscriptstyle \Lambda_2} && \supset \mathcal{L}_{\scriptscriptstyle \text{CFT}_{\rm D}} + g_{\scriptscriptstyle \rm D} A_{{\scriptscriptstyle \rm D} \mu} J_{\scriptscriptstyle \rm D}^\mu + \frac{1}{\Lambda_1} \rho_{B \mu\nu} \bar{\psi}_L \sigma^{\mu\nu} \chi_R + \frac{1}{\Lambda_2} F_{{\scriptscriptstyle \rm D} \mu\nu} \bar{\psi}_{\scriptscriptstyle \rm comp,R} \sigma^{\mu\nu} \chi_{\scriptscriptstyle \rm comp, L} \label{eq:Theory_at_Lambda_2} \\
&& + \tilde{y}_{\Psi} \left( \frac{\Lambda_2}{\Lambda_1} \right)^{d_R-5/2} \Lambda_2  \sum_n c_n \bar{\psi}_L \psi_{\scriptscriptstyle \rm comp, R}^{(n)} + \tilde{y}_{\chi} \left( \frac{\Lambda_2}{\Lambda_1} \right)^{d_L-5/2} \Lambda_2  \sum_m d_m \bar{\chi}_{\scriptscriptstyle \rm comp, L}^{(m)} \chi_R. \nonumber
\eea
\begin{figure}[h]
  \centering
  \includegraphics[width=0.70\textwidth]{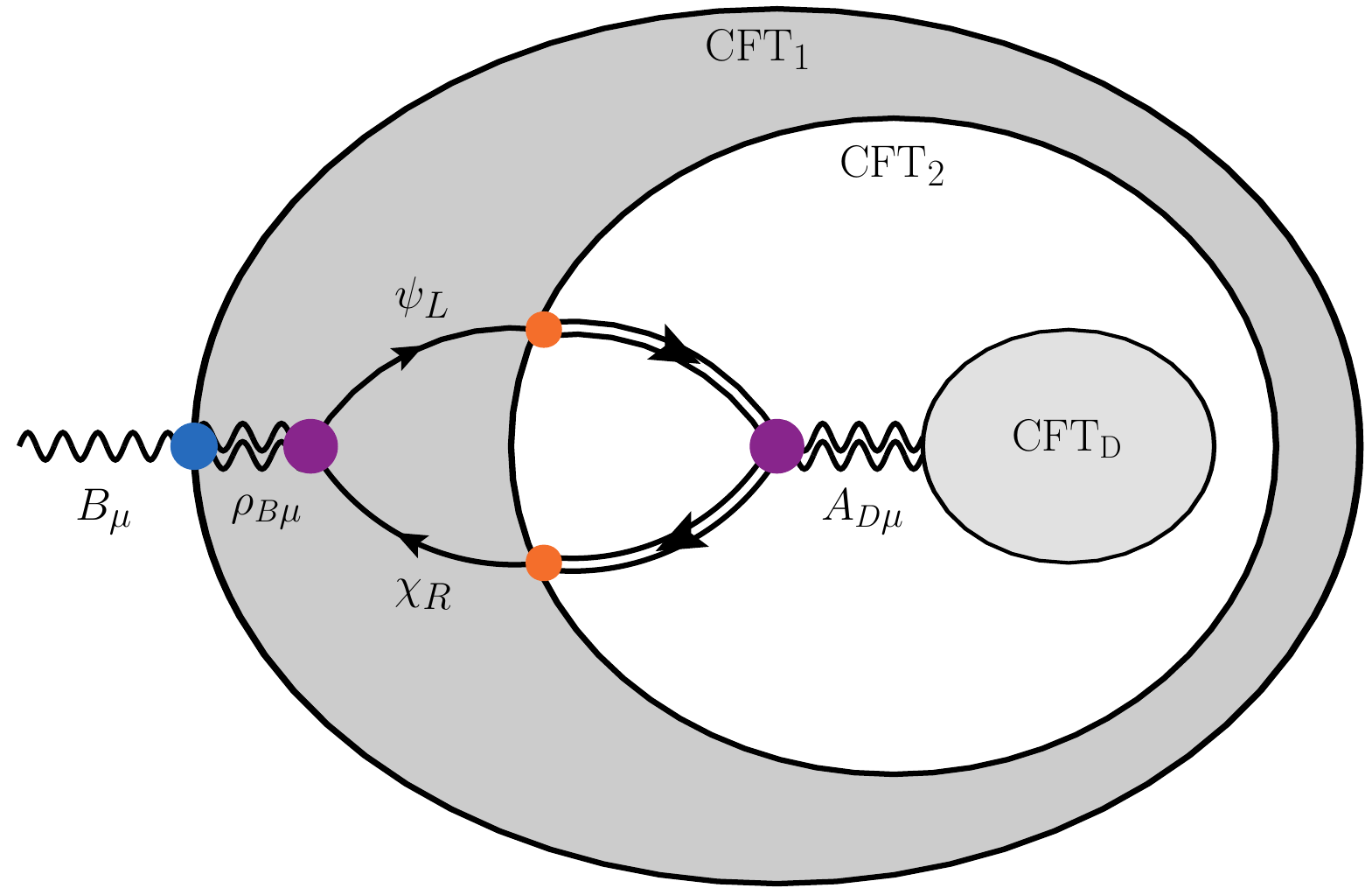}
  \caption{A diagram responsible for the effective mixing between $B_\mu$ and $A_{{\scriptscriptstyle \rm D} \mu}$. $B_\mu$ is an elementary gauge boson external to $\text{CFT}_1$ and $\rho_{B\mu}$ is a composite vector meson of confined phase of $\text{CFT}_1$. This latter phase also includes a pair of composite fermions denoted as $\psi_L$ and $\chi_R$ which couples to a composite $\text{CFT}_2$ as described in \autoref{eq:Theory_at_Lambda_1}. These couplings lead to partial-compositeness coupling once $\text{CFT}_2$ confines. This is shown as orange blobs. The confined $\text{CFT}_2$ also contains a composite vector meson $A_{{\scriptscriptstyle \rm D} \mu}$ which itself is external to $\text{CFT}_{\rm D}$.   }
  \label{Fig:CFT_diagram_effective_mixing}
\end{figure}

From this, we can estimate the effective kinetic mixing between the elementary $U(1)_Y$ gauge boson $B_\mu$ and $A_{{\scriptscriptstyle \rm D} \mu}$ by evaluating the diagram shown in \autoref{Fig:CFT_diagram_effective_mixing}. From \autoref{Fig:CFT_diagram_effective_mixing} it is clear that the effective mixing is tiny due to two factors of fermion mixing since the latter two are very small by a RG evolution. In the end, we get
\beq
\tilde{\epsilon} \sim \frac{g}{g_{1 s}} \frac{\tilde{y}_{\Psi} \tilde{y}_{\chi}}{16\pi^2} \left( \frac{\Lambda_2}{\Lambda_1} \right)^{d_L+d_R-4}.
\label{eq:effective_mixing}
\eeq
The factor $g/g_{1 s}$ is from the elementary-composite mixing between $B_\mu$ and $\rho_{B\mu}$ as explained above, and we note that this estimation is up to possible $\log \left(\Lambda_2 / \Lambda_1 \right)$. We recall that $d_L + d_R > 10$ and therefore, $\tilde{\epsilon}$ can be highly suppressed by virtue of the RG running factor. 

Since the dark CFT is uncharged under $B_\mu$, the leading order interaction is expected to be the dipole-type. The interaction strength can be estimated to be (dropping the subscripts, e.g.~$\Lambda_2 \to \Lambda$, to get an expression used in the rest of the paper)
\beq
\mathcal{L}_{\rm \scriptscriptstyle COFI} \supset \frac{\lambda}{\Lambda^{d-2}} B_{\mu\nu} \mathcal{O}^{\mu\nu}.
\label{eq:Theory_COFI}
\eeq
where $d$ is the scaling dimension of $\mathcal{O}^{\mu\nu}$ and $\lambda \sim g_{\scriptscriptstyle \rm D} \tilde{\epsilon}$ and thus can be readily very small.\footnote{The superficial IR-divergence from the intermediate $A_{{\scriptscriptstyle \rm D} \mu}$ propagator is absent thanks to two-derivatives from $\rho_{B \mu\nu} F_{\scriptscriptstyle \rm D}^{\mu\nu}$.}

Finally, we show that asymmetric reheating requires $T_R < \Lambda_2$. Suppose that the decay of the inflaton reheats the \emph{external} sector plasma to a temperature $\Lambda_2 < T_R \lesssim \Lambda_1 $. This means that the correct description of the theory right after reheating is that of \autoref{eq:Theory_at_Lambda_1}. This comes with sizable coupling between the CHM sector and $\text{CFT}_2$. For a generic CFT, the entirety of $\text{CFT}_2$ will then be thermalized via this coupling. In particular, it is unlikely that there is a subsector of $\text{CFT}_2$ which is isolated and remains ``cold''. Once the universe cools to $T \sim \Lambda_2$, $\text{CFT}_2$ confines and, in particular, a \emph{thermal} $\text{CFT}_{\scriptscriptstyle \rm D}$ appears. So for a generic $\text{CFT}_2$, $\text{CFT}_{\scriptscriptstyle \rm D}$ will be at roughly the same temperature as the SM sector. On the other hand, if $T_R < \Lambda_2$, then the right description after reheating is \autoref{eq:Theory_COFI}, which comes with highly suppressed coupling.

\subsubsection*{(ii) $d_L, d_R < 5/2$}

We briefly discuss the case with $d_L, d_R < 5/2$. The other choices of $d_L$ and $d_R$ are then simply mixture of the two cases we describe. 

When $d<5/2$, the linear mixing in \autoref{eq:Theory_at_Lambda_1} is a relevant operator and grows in the IR. The RG running is described by (see e.g.~\cite{Contino:2010rs, Agashe:2015izu})
\beq
\mu \frac{d y}{d\mu} = \gamma y + c \frac{N}{16\pi^2} y^3 + \cdots
\eeq
where $\gamma = d - 5/2 < 0$ is the anomalous dimension of the CFT operator $\mathcal{O}$,  $N$ denotes the number of ``color'' of the gauge theory describing the CFT, and $c$ is an ${\mathcal{O}}(1)$ number. RG flow increases $y$ and at some point the second term becomes as important as the first. Provided $c>0$, there exists an IR fixed point where $y$ stops running. We name the scale of the fixed point $\Lambda_*$ and the coupling at the fixed point $y_* \sim \tilde{y} (\Lambda_* / \Lambda_1)^{d-5/2}$, which can be $\mathcal{O} (1)$. At the fixed point, the linear mixing terms become marginal operators and, at the same time, the fermion fields $\psi_L$ and $\chi_R$ acquire sizable anomalous dimension. Explicitly, the scaling dimensions of them become $[ \psi_L ] = 4-d_R > 3/2, \; [ \chi_R ] = 4-d_L > 3/2$. 

Unlike in the first case with $d_L, d_R > 5/2$, the fermion mixings are sizable and one may conclude that the effective mixing $\tilde{\epsilon}$ is not suppressed anymore. This, however, is not true. The anomalous dimensions of $\psi_L$ and $\chi_R$ make the dipole interaction appearing in \autoref{eq:Theory_at_Lambda_1} very irrelevant interaction with scaling dimension $10-d_L-d_R > 5$. Via RG evolution, this means that this dipole operator becomes highly suppressed at $\Lambda_2$.
\beq
\frac{1}{\Lambda_1} \left( \frac{\Lambda_2}{\Lambda_*} \right)^{5-d_L-d_R} \Lambda_2^{d_L+d_R-5} \rho_{B \mu\nu} \bar{\psi}_L \sigma^{\mu\nu} \chi_R
\eeq
It is then straightforward to estimate the effective mixing $\tilde{\epsilon}$. The final result is in fact the same as \autoref{eq:effective_mixing}. Interestingly, despite having very different RG evolutions, the product of the dipole interaction and the fermion mixings appearing in \autoref{Fig:CFT_diagram_effective_mixing} stays the same in both cases. A similar phenomenon appeared in the neutrino mass from a warped 5d model (and its 4d CFT dual) \cite{Agashe:2015izu}.

\subsubsection{Summary for the UV theory}

To sum up, asymmetric reheating is achieved by virtue of composite$\text{--}$elementary division\footnote{It is this composite$\text{--}$elementary division that distinguishes our theory from the UV completion of COFI by a weakly coupled gauge theory with a IR fixed point proposed in \cite{Hong:2019nwd}. In the latter case, unless symmetry forbids, generically there will be couplings between the gauge theory sector and the inflaton, and in turn the dark CFT sector will inherit a unsuppressed coupling to the inflaton. } and the dynamically generated small coupling, provided the reheat temperature satisfies $T_R < \Lambda_2$. Specifically, the composite$\text{--}$elementary division makes it natural that the primordial reheating occurs only for the external states, hence only the SM sector. Then, the small coupling between the SM and dark CFT sectors, induced by RG running followed by a confining phase transition, forbids an efficient energy transfer from the SM to the dark CFT.

\subsection{IR effective theory, mass gap, and composite dark photon}
\label{subsec:IR EFT}

In this section, starting from \autoref{eq:Theory_COFI}, we explain the mass gap generation, IR effective theory below the mass gap $m_{\rm gap}$ and comment on notable features of our model. 

We first note that since our model is based on a tensor operator, the RG running of the CFT and dynamical mass scale generation do not go through the mechanisms introduced in \cite{Hong:2019nwd,Hong:2022gzo}. In particular, the operator mixing effects \cite{Hong:2022gzo} which makes the COFI-mechanism generic for the case of scalar operator do not occur in our model. Instead, a necessary scalar deformation may arise from the operator product expansion (OPE) $\mathcal{O}_{\mu\nu} \times \mathcal{O}^{\mu\nu} $. We discuss this in detail in \autoref{sec:dsScales}. Here, we simply assume that such a scalar CFT operator exists and explore its implications on the IR EFT. 

If such a deformation is close to being marginal, the theory described by \autoref{eq:Theory_COFI} undergoes a slow RG running (walking). At $E \sim m_{\rm gap}$, the conformal invariance is spontaneously broken and a gap scale is generated. By virtue of walking, the separation between $\Lambda_2$ and $m_\text{gap}$ are generically large.
 
We make a simplifying assumption that the spontaneous conformal symmetry breaking is a confining phase transition and a spectrum of composite hadrons become the relevant degrees of freedom in the IR. The operator $\mathcal{O}^{\mu\nu}$ is then interpolated by\footnote{In principle, the operator can have a non-zero overlap with a composite 2-form field $C_{\mu\nu}$. For our purposes, it suffices to assume that $\mathcal{O}_{\mu\nu}$ has unsuppressed overlap with kinetic term of composite dark photon.}
\beq
\mathcal{O}^{\mu\nu} \sim \frac{1}{g_s} m_{\rm gap}^{d-2} \rho_{\mu\nu},
\label{eq:dark photon from O}
\eeq
where $\rho_{\mu\nu}$ is the field strength of the composite vector meson $\rho^\mu$; the dark photon in our theory\footnote{Strictly speaking, the dark photon is a mixture of $A_D$ and $\rho$, but it will be mostly comprised of $\rho$}.  The dependence on $m_{\rm gap}$ is fixed by dimensional analysis and $g_s \sim \frac{4\pi}{\sqrt{N}}$ is the coupling constant among the composite states, where $N$ is the number of ``color'' of the gauge group in the CFT.

The confined phase of the dark CFT may contain a (or a set of) Goldstone boson $\pi$ and they can play the role of dark matter in our theory. Using \autoref{eq:dark photon from O}, we obtain the IR effective theory of hadrons from \autoref{eq:Theory_COFI}
\bea
&& \mathcal{L}_{\rm \scriptscriptstyle IR} \sim \frac{1}{2 g_s^2} \rho_{\mu\nu} \rho^{\mu\nu} + \epsilon B_{\mu\nu} \rho^{\mu\nu} + \partial_\mu \pi^+ \partial^\mu \pi^- + m_\text{DM} \pi^+ \pi^- + ig_s \rho^\mu \pi^+\overset{\leftrightarrow}{\partial}_\mu \pi^-, 
\label{eq:IR_EFT}
\eea
with the kinetic mixing given by 
\bea
 \epsilon = \frac{\lambda}{g_s} \left( \frac{m_{\rm gap}}{\Lambda} \right)^{d-2}. 
 \label{eq:kmixing}
\eea
We will now make a few comments on the low energy effective theory. The effective kinetic mixing $\epsilon$ shown in \autoref{eq:kmixing} is naturally small. In particular, in addition to the small $\lambda$ (whose natural smallness was explained earlier), it gets further suppressed by the RG running factor $\left( \frac{m_{\rm gap}}{\Lambda} \right)^{d-2}$ (recall $d \geq 2$ by unitarity). This latter factor exhibits the interesting fact that a smaller $m_{\rm gap}$ implies a smaller mixing $\epsilon$. This has a straightforward physics interpretation. We first note that $m_{\rm gap}$ is a consequence of conformal invariance breaking effect, thus the size of $m_{\rm gap}$ is positively correlated with the size of the breaking. In general, a smaller $m_{\rm gap}$ means slower (hence longer) RG running of the CFT sector. On the other hand, the mixing, $\epsilon$, is induced from the coupling between the SM and CFT sectors in the UV theory. The unitarity bound $d \geq 2$ implies that the interaction \autoref{eq:Theory_COFI} is an irrelevant operator. This in turn suggests that smaller $m_{\rm gap}$ results in larger suppression of $\epsilon$ from longer RG running. 

In addition to the kinetic mixing between the dark photon and hyper-charge gauge boson, we included terms for the dark matter candidate, $\pi^{\pm}$, and their interaction with the dark photon. Here, we assume that dark matter particles are pseudo-Nambu-Goldstone bosons (pNGB) of the spontaneously broken global symmetry of the CFT. Their mass is controlled by the size of the explicit breaking of the global symmetry, which we take to be a free parameter.  The ratio $r\equiv m_\text{DM} / m_{\rm gap}$ can be smaller than one, which ensures that dark matter can easily be lightest stable particle\footnote{From the form of the effective Lagrangian, we've implicitly assumed that $\pi$ has a dark charge which ensures stability. If we assumed no such dark charge, then one might expect $\mathcal{O}^{\mu\nu}$ can also interpolate to an operator of the form $\sim\pi\rho^{\mu\nu}$. After kinetic mixing, this will allow the process $\pi\rightarrow\gamma\gamma$. However, the LO decay rate will go as $\epsilon^4$; ensuring that its cosmologically long-lived.}. 

Dark matter will couple to the dark photon in the same manner as the pions in low energy QCD interact with the $\rho$-meson. The strength of the coupling is $g_s \sim \frac{4\pi}{\sqrt{N}}$. For a reasonable choice of $N$ consistent with large-$N$ treatment, we can take $g_s \sim \mathcal{O} (1)$. This coupling induces self-interaction among dark matter states. For $m_{\rm gap} \lesssim \mathcal{O} (100) {\rm MeV}$ there are non-trivial constraints on this DM self-interaction, e.g.~from observation of the bullet cluster. As we discuss later in \autoref{subsec:constraints} (also discussed in \cite{Hong:2019nwd, Hong:2022gzo}), this constraint can be avoided with a proper choice of the ratio $ m_\text{DM} / m_{\rm gap} < 1$. Furthermore, any relevant processes involving the visible sector and the dark matter candidate is independent of $g_s$.

Other than the DM particles, the rest of the hadrons in the confined CFT are expected to have mass on the order of $m_{\rm gap}$, which we assume to hold for our model. In particular, the dark photon, as one of the normal composite states, is assumed to have mass $\sim m_{\rm gap}$. We suppress the rest of the hadrons from our effective theory.

\section{Dark Matter Phenomenology}
\label{sec:Pheno}

In this section, we describe the dark matter phenomenology of our theory.
In \autoref{subsec:relic density} we discuss the cosmological evolution of the energy density in the dark sector. In \autoref{subsec:ir-dom}, we present a parameter scan which reproduces the observed relic density for IR-dominant production and discuss the main characteristics. In \autoref{subsec:UV-dominant}, we present a parameter-scan for UV-dominant production and discuss the associated physics.  Lastly, in \autoref{subsec:constraints}, we discuss theoretical constraints and relevant observational constraints, including DM self-interaction, warm DM bound, star cooling bound and more. Throughout this section, in order to avoid interrupting the flow of the discussion, we relegate technical details to several appendices (see \autoref{app:COFI_computation} and \autoref{app:star_cooling}).

\subsection{Dark matter production mechanisms}
\label{subsec:relic density}

The details of the freeze-in production of dark matter in this model depend on the nature of the coupling in \autoref{eq:Theory_COFI}, especially the scaling dimension $d$ of the operator $ \mathcal{O}^{\mu\nu}$. At the same time, it also depend on various scales in the problem, including the temperature in the SM sector $T$, the temperature in the dark sector $T_{\rm ds}$, $m_{\rm gap}$, and the dark matter mass $m_{\rm DM}$. 

If the temperature of the SM sector $T$ is larger than $m_{\rm gap}$, then the freeze-in processes produce CFT objects in the final state. We denote this as COFI production. In the other regime, $T < m_{\rm gap}$, the final state consists of the ``hadronic'' states of the confined CFT and the physics becomes that of standard particle production. 

The dark sector is assumed to thermalize with itself.\footnote{This condition can easily be satisfied in large-$N$ CFT. Specifically, since we consider non-thermal freeze-in production, our conformal dark sector is a CFT at very low temperature which is strongly interacting. This also allows us to use AdS/CFT duality.} When the dark sector is radiation-like, its temperature, $T_\text{ds}$, is given by 

\begin{equation}
  \rho_\text{ds}=AT_\text{ds}^4,
\end{equation}
where  $A$ is the analog of $\frac{\pi^2}{30} \times {\rm dof}$ appearing in the energy density of a relativistic fluid.

As shown in \cite{Hong:2019nwd}, IR-dominant COFI production can occur if the sum of the scaling dimensions of the operators appearing in the interaction term \autoref{eq:Theory_COFI} is less than or equal to 9/2. In our case, this requires $2< d < 5/2$, where the lower limit is the unitarity bound. However, as we will show, this conclusion is based on the assumption that the dark sector is thermalized to a temperature $T_\text{ds} > m_\text{DM}$ during COFI production. In order to clarify this point, let us first briefly review the COFI production, obtain the bound $d < 5/2$, and generalize it to the case $T_\text{ds} < m_\text{DM}$. 

\subsubsection{$T_\text{ds} > m_\text{DM}$ during the COFI production}
\label{subsubsec:COFI_R}

Starting from the general Boltzmann equation (BE), the relevant equation for COFI is (see \cite{Hong:2019nwd} for details)
\beq
  \frac{d}{dt}\rho_\text{ds}+3H(\rho_\text{ds}+P_\text{ds})=\Gamma,
\label{eq:BE_R}
\eeq
where $\Gamma$ is the energy transfer rate per volume from SM to CFT. We've dropped the energy transfer from the CFT sector to the SM sector due to the assumption that the CFT energy density is always small compared to the SM.
In our case, $\Gamma$ takes a general form
\beq
\Gamma = \left(\frac{\lambda}{\Lambda^{d-2}} \right)^2 \Gamma_d T^{2d+1}
\label{eq:Collision term general form}
\eeq
with a process- and $d$-dependent coefficient $\Gamma_d$. In order to simplify expressions, we further define $B_d = \left( \frac{\lambda}{\Lambda^{d-2}} \right)^2 \Gamma_d$. For our model, the population of the dark sector can occur via $f \bar{f} \to B_\mu \to {\rm CFT}$: annihilation of SM fermion pairs through the exchange of the hypercharge gauge boson. In addition, at finite temperature, the photon acquires a thermal mass $m_p$. Following \cite{Dvorkin:2019zdi}, we take $m_p$ to be roughly the plasma frequency\footnote{The effective in-medium mass is generically a function of the momentum and the polarization mode.}
\begin{equation}
  m_{p} \approx \omega_p\approx \frac{eT}{3} \approx 0.1 T,
\end{equation}
where $e$ is the electric charge. The plasmon can directly decay into the CFT state which contributes to the production. At $T > v$, the intermediate state in the fermion annihilation is the $U(1)_Y$ gauge boson. Below $v$, it becomes a linear combination of the photon and $Z$ gauge boson. At $T > m_{\rm gap}$ the final state is the CFT state, while for $T < m_{\rm gap}$ it is the hadronic state of the confined CFT.\footnote{There is, in principle, contribution from pair annihilation of the Higgs doublet (equivalent to $Zh$ annihilation below EWSB). In the case of IR freeze-in, since this process shuts off at scales well above the dark matter mass that we are considering. Its contribution to the overall relic density is negligible. In the case of UV freeze-in, due to the large number of fermions charged under $U(1)_Y$, its contribution is subleading compared to the fermion annihilation process.} 

The collision terms take the general form \autoref{eq:Collision term general form} and as we show in detail in \autoref{subapp:fermion_pair_ann} the coefficients $B_d$ are given by 
\bea
&& B_d (\bar{f} f \to \text{\tiny CFT}) = \frac{12 d}{(2\pi)^{2d+2}} \left( \frac{\lambda e}{\Lambda^{d-2}} \right)^2,
\label{eq:B_d_ff_ann} \\
&& B_d (\gamma^* \to \text{\tiny CFT}) = \left( \frac{6 A_d e^{2d-4}}{3^{2d-4} \pi^2} \right) \left( \frac{\lambda e}{\Lambda^{d-2}} \right)^2,
\label{eq:B_d_photon_decay}
\eea
where $A_d$ is related to the phase space of CFT state and is defined in \autoref{eq:A_d}.

Now, let's move onto the LHS of the Boltzmann equation. Rotational and conformal invariance (implying $T^{\mu\nu}$ is traceless) tells us that $P_\text{ds}=\frac{1}{3}\rho_\text{ds}$. Here, it is important to realize that usage of this dispersion relation is valid only if $T_\text{ds} > m_\text{DM}$.\footnote{Strictly speaking, the dark sector plasma is that of a CFT (as opposed to the ``hadronic'' phase) only if $T_\text{ds} > m_{\rm gap}$. However, here we are using the fact that so long as $T_\text{ds} > m_\text{DM}$ \emph{and} if most of the energy density of the dark sector is rapidly transferred to dark matter state, then the energy density behaves as a relativistic gas.} With that, we have
\beq
  -HT \rho_{\rm ds}'+4H\rho_{\rm ds}=B_d T^{2d+1},
  \label{eq:BE}
\eeq
where $'$ is a derivative with respect to $T$. In the radiation dominated epoch,
\beq
  H=\sqrt{g_*}\frac{T^2}{\text{m}_\text{pl}}, \;\; \text{m}_\text{pl}\equiv \frac{3\sqrt{5}}{2\pi^{3/2}}M_\text{pl} \approx 7.35 \times 10^{18} {\rm GeV}.
\eeq
Ignoring the temperature dependence of the number of relativistic degrees of freedom $g_*$\footnote{This is a simplification made here for illustrative purpose. In our numerical results, we include the effect of time dependence of $g_*$}, the solution to  \autoref{eq:BE} is of the form 
\beq
  \rho_\text{ds} (T) = \frac{B_d \text{m}_\text{pl}}{\sqrt{g_*}(5-2d)} T^4 \left( T^{2d-5} - T_R^{2d-5} \right), 
\label{eq:rho_CFT_relativistic}
\eeq
where $T_R$ is the reheat temperature (we take $T_R \sim \mathcal{O} ({\rm TeV})$) and we have used the initial condition $\rho_\text{ds}(T_R)=0$. We factored out an overall factor of $T^4$ which allows us to interpret the expression in the parenthesis as the change of energy density in the comoving frame.
 
Whenever the production (for each channel) ends at $T$ sufficiently lower than $T_R$, for $d<5/2$, we can safely drop the $T_R$-dependent term. This shows that the production is insensitive to the UV physics (i.e. IR-dominant). Conversely, when $d>5/2$, the $T$-terms gets dropped. This demonstrates that the production is only sensitive to the UV physics (i.e. UV-dominant).

\subsubsection{$T_\text{ds} < m_\text{DM}$ during the COFI production}
\label{subsubsec:COFI_NR}

If $T_\text{ds}$ drops below $m_\text{DM}$ during the production, then the equation of state changes to $P_{\rm ds}=0$. 
The subsequent evolution of the energy density obeys
\beq
  -HT\rho_{\rm ds}'+3H\rho_{\rm ds}=B_d T^{2d+1}.
\label{eq:BE_NR}
\eeq
Here, we encounter another important temperature threshold, $T_{\rm NR}$.  This is the temperature of the SM bath when the dark sector temperature drops to $m_\text{DM}$. After this point, all particle states in the dark sector are non-relativistic. 

The solution to the Boltzmann equation for temperatures below $T_\text{NR}$ is then given by 
\beq
  \left.\rho_\text{ds} (T)\right|_{T<T_\text{NR}} =\frac{B_d \text{m}_\text{pl}}{\sqrt{g_*}(4-2d)} T^3 \left( T^{2d-4}  - T_{\rm NR}^{2d-4} \right) + \left( \frac{T}{T_{\rm NR}} \right)^3 \rho_{\rm ds} (T_{\rm NR}) ,
  \label{eq:rho_CFT_NR}
\eeq
where again we pulled out the overall factor $T^3$ (the appropriate scaling for matter-like energy density). The second term is simply the evolution of the energy density produced prior to reach this point. 
As before, the expression in the parenthesis in the first term is the change of energy density in the comoving frame. Crucially, for all $d \geq 2$ (which is always the case for the interaction in \autoref{eq:Theory_COFI}), the production is UV-sensitive. Hence, reaching $T_{\rm NR}$ provides an effective endpoint to COFI production. 

For the special case where the dark sector was never relativistic, this corresponds to setting $T_{\rm NR} = T_R$ and $\rho_{\rm ds} (T_R) = 0$. 
In doing so, one can see that the late time energy density only depends on $d$, $\lambda$, $T_R$, and $\Lambda$ (provided that the IR scale is much smaller than $T_R$).

\subsubsection{``Hadronic'' production}

If production continues to occur when the temperature of the SM bath is less than $m_{\rm gap}$, the characteristic energy of the initial SM states is also less than the mass gap. In this case, the produced final states are ``dark hadrons'' rather than CFT states\footnote{This ``hadronic'' production is strictly speaking not a conformal freeze-in and instead is the usual particle freeze-in.}. While in principle, the relevant processes and their rates are model-dependent, in the region where $m_\text{DM}$ is modestly smaller than $m_{\rm gap}$\footnote{This will turn out to be a necessary condition in order to evade the DM self-interaction bound (see \autoref{subsubsec:DM_SI}).} we obtain a reasonably reliable and simple description as follows. As we show in detail in \autoref{sec:HadProduction}, the ``hadronic'' production process is (i) UV-dominant (i.e.~most of the production occurs at $T \sim m_{\rm gap}$) and is (ii) subdominant to the energy injected via COFI established at $T > m_{\rm gap}$ (if at all). 

\subsubsection{Post-production evolution}
\label{subsubsec:post-prod}

Freeze-in production is terminated by a threshold effect. This can be a result of switching to non-relativistic production, switching to the ``hadronic'' production mode, or the initial states decoupling from the SM bath.

Let $T_f$ be the  threshold that puts an end to the production. The subsequent evolution depends on whether the dark sector is radiation-like or matter-like. If it were radiation-like, we get today's dark matter energy density, $\rho_{\rm ds, 0}$, by first redshifting as radiation (i.e. as $T^4$) down to $T_\text{NR}$ and further redshifting the energy density as matter (i.e.~$T^3$) between $T_{\rm NR}$ and today:
\beq
\rho_{\rm ds, 0} = \rho_{\rm ds}(T_f) \left( \frac{T_{\rm NR}}{T_f} \right)^4 \left( \frac{T_0}{T_{\rm NR}} \right)^3=Am_\text{DM}^4\left( \frac{T_0}{T_{\rm NR}} \right)^3.\label{eq:postprod}
\eeq

\subsection{IR-dominant freeze-in}

Based on the discussion in \autoref{subsec:relic density}, we are in a position to compute relic abundances of dark matter and study its dependence on various parameters in this model. As is clear from \autoref{eq:Theory_COFI}, the physics of COFI is controlled mainly by two parameters, $\lambda$ and the scaling dimension $d$ of the CFT operator. These can be traded with $m_{\rm gap}$ and $d$ (see the discussion around \autoref{eq:Os_OPE_1} and \autoref{eq:CFT_deformation}). These parameters will be scanned over in our plots. The remaining model parameters: $T_R$, $\Lambda$, $r$, and $A$, are fixed for the plots. We are mainly interested in $T_R \approx \Lambda \sim \mathcal{O} (\text{TeV})$; with $T_R<\Lambda$ to ensure the validity of the theory throughout the entire freeze-in process and asymmetric reheating. The dependence on $A$ is pretty mild. We consider two choices for $r$: $r=0.1$ and $r=0.01$. Finally, depending on scaling dimension $d$, we have two qualitatively different scenarios. We begin with the so called IR-dominant case, with $2<d<2.5$, leaving the UR-dominant production to the next subsection. 

\label{subsec:ir-dom}
\begin{figure}
  \centering
  \includegraphics[width=0.47\textwidth]{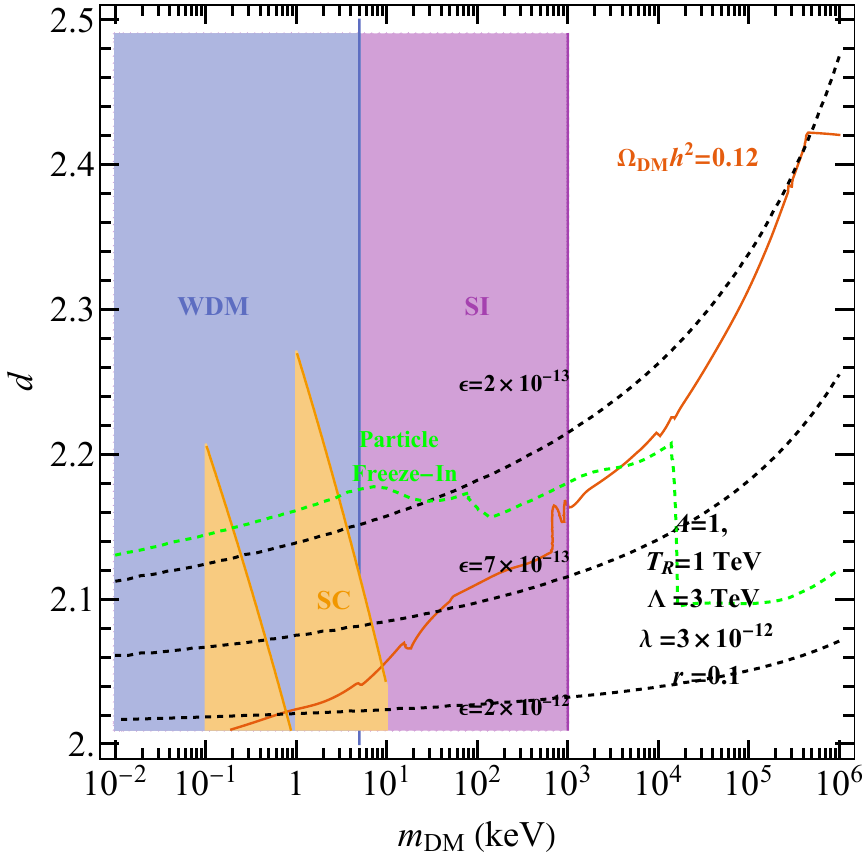}
  \includegraphics[width=0.47\textwidth]{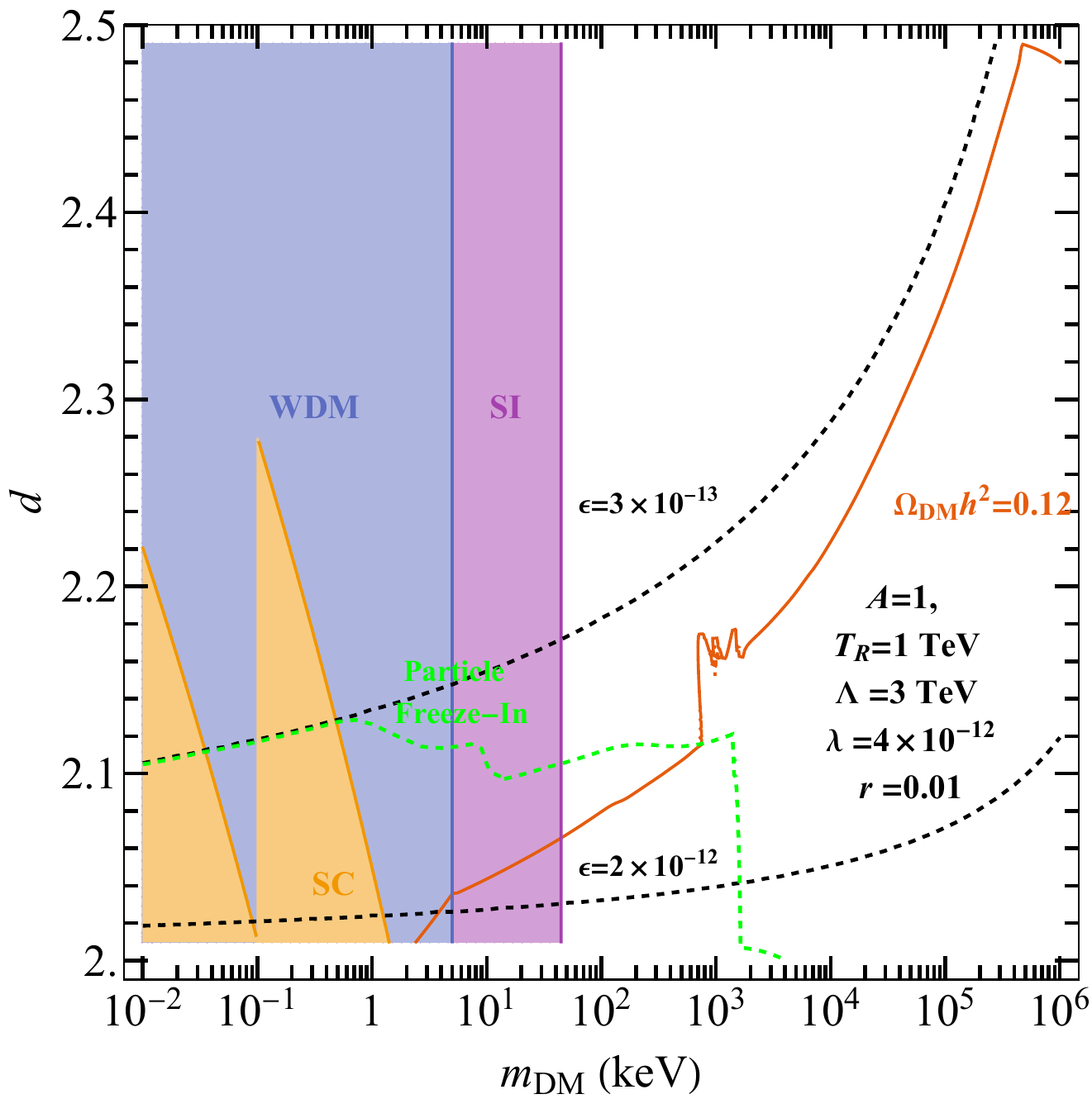}
  \caption{The dark matter mass and the CFT operator scaling dimension which reproduces the dark matter relic density (orange-red solid line) with $r=0.1$ (left) and $r=0.01$ (right). For comparison, the green dashed line shows an estimate for the expected parameters which reproduces the observed relic density for the ``usual'' freeze-in with kinetic mixing parameter given by \autoref{eq:kmixing}  and the same relation between the dark photon mass and dark matter mass \cite{Blennow:2013jba}. The blue shaded region corresponds to the region of parameter space excluded by the warm dark matter bound. The yellow shaded region corresponds to the stellar cooling bound. The purple shaded region is excluded by the DM self-interaction bound coming from the observation of bullet-cluster. The dashed curves show contours of constant kinetic-mixing parameter, $\epsilon$, with $g_s=1$. }
  \label{Fig:relic_density_constraints_IR_dom}
\end{figure} 

The contours of the observed relic density in the plane of $( m_\text{DM}, d )$ (with the remaining parameters fixed) is shown in orange-red in \autoref{Fig:relic_density_constraints_IR_dom}. For both panels, there is a general tendency for $d$ to increase with $m_\text{DM}$. The physics behind this is that as $m_\text{DM}$ increases, so does $T_{\rm NR}$ 
This in turn means that $\rho_{\rm ds}$ starts redshifting as matter at a higher temperature; leading to an effective increase of the final relic density, $\rho_{\rm ds, 0}$. This increase must be compensated for by adjusting $d$ so that $\rho_{\rm ds, 0}$ matches a constant observed value. A larger $d$ corresponds to a more irrelevant interaction, hence slower heating. Therefore, an increase in $m_\text{DM}$ is generically balanced by an increase in $d$, as observed in \autoref{Fig:relic_density_constraints_IR_dom}. Following this discussion, one can also determine how the general trend of the contour will behave as we lower our choice $\lambda$ (or equivalently raising $\Lambda$). The scattering process will inject less energy into the dark sector. To compensate, $m_\text{DM}$ can be raised so the dark sector can start redshifting as matter earlier or $d$ can be lowered to make the interaction more relevant. As such, the contour will shift towards the bottom right.

\begin{figure}
  \centering
  \includegraphics[width=0.9\textwidth]{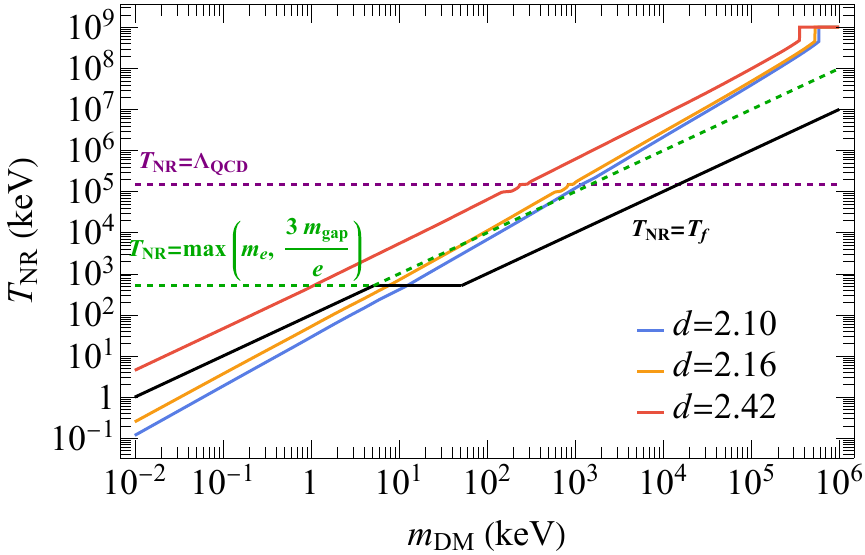}
  \caption{$T_{\rm NR}$ as a function of $m_\text{DM}$ with a couple of choices of $d$ for the parameters chosen in the left panel of \autoref{Fig:relic_density_constraints_IR_dom} (i.e. $r=0.1$). The green-dashed curve shows the temperature when one of the two SM initial states decouple from the thermal bath. The black solid line shows the temperature when either all of the SM initial states decouple or when we switch to ``hadronic'' production. For data points below the black solid line, the transition into the non-relativistic phase occurs after the end of production.}
  \label{Fig:T_NR}
\end{figure}

Both plots exhibits some sharp change of slope for large $m_\text{DM}$. In the case of the left plot, it plateaus; whereas the right plot drops sharply. Both of these correspond to the case with $T_\text{NR}=T_R$. This can be verified by following the red curve in \autoref{Fig:T_NR}. As discussed in \autoref{subsubsec:COFI_NR}, this results in the late time energy density becoming independent of $m_\text{DM}$. The discrepancy in the $r=0.01$ plot is due to the fact that $T_\text{SM}=m_\text{gap}$ and $m_p(T_\text{SM})=m_\text{gap}$ occur at scales close to $T_R$. When this happens, the first term in \autoref{eq:rho_CFT_NR} cannot be ignored and the interplay between the two terms determines the overall shape.

A sharp change of slope also occurs for low $m_\text{DM}$: at $m_\text{DM}\approx 50~\text{keV}$ for the plot and $m_\text{DM}\approx 5~\text{keV}$ for the right figure. This is a result of $T_\text{NR}$ dropping below the other possible endpoints of production. This can be verified by checking that the intersection of the blue curve with the black curve in \autoref{Fig:T_NR} does indeed occur at $m_\text{DM}\approx 50~\text{keV}$ for $r=0.1$.

There are several localized bumps and dips in \autoref{Fig:relic_density_constraints_IR_dom}. They arise from jumps in the number of relativistic degree of freedom, $g_*$, and the number of production channel. We explain this focusing on the left panel (with $r = m_\text{DM}/m_{\rm gap} = 0.1$), but our discussion applies in general. 

For example, there are noticeable bumps at $m_\text{DM} \approx$ MeV, $d\approx2.16$. These features are related to $T_{\rm NR}$ crossing some mass threshold:  $\Lambda_{\rm QCD}$ followed by $m_\mu$ for the subsequent bump. This fact can be checked by finding the intersection of the yellow curve with the purple dashed line in \autoref{Fig:T_NR}. 

To understand this better, we note that $T_{\rm NR}$ decreases as we lower $m_\text{DM}$. When $T_\text{NR}$ happens to cross a threshold, e.g.~the electron mass, there are potentially two effects. First, it reduces the number of production channels, effectively decreasing $\rho_{\rm ds, 0}$. This must be compensated for by decreasing $d$ which has an effect of increasing the rate of heating, and hence the final energy density. The second effect is the change of $g_*$. We incorporated the change of $g_*$ numerically by evaluating the energy density exactly outside of any phase transitions, hence the ``jump'' in $g_*$ across each mass threshold (except the QCD phase transition and neutrino decoupling) is rather smoothed out. The (smooth) decrease in $g_*$ results in increase in $\rho_{\rm ds, 0}$, which then needs to be balanced by increasing $d$. This explains the smooth rising section right after the drop. 

During the QCD phase transition, $g_*$ drops sharply (increase in $\rho_{\rm ds, 0}$) and  the up, down, and strange quarks decouple (decreasing the production channels, hence $\rho_{\rm ds, 0}$). Numerically, it turns out that the former effect dominates and is compensated by a sharp increase in $d$. Soon after, the muon decouples. This time, the effect of decreasing the number of production channels is larger; requiring a decrease in $d$. 

While the qualitative feature described above is solid, the details of the shape appearing in \autoref{Fig:relic_density_constraints_IR_dom} is partly due to the way we implemented $g_*$ and changes in the production channel. Furthermore, the impacts of neglecting the derivative of $g_*$ in the BE can be large. As such, the shape of the contours in that region should not be taken to be exact.

For comparison, we have also drawn an estimate for the relic density contour for the particle freeze-in scenario based on the results given in \cite{Blennow:2013jba} in green\footnote{The exclusion contours shown in the figure do not apply to this contour.}. To obtain this curve, we assumed that the kinetic mixing parameter is given by \autoref{eq:kmixing} and the dark photon mass obeys $rm_{A'}=m_\text{DM}$. Using the right plot as an example, for low dark matter masses, we see that the contour exactly follows the contour of constant kinetic mixing parameter. As $m_\text{DM}$ increases, the contour is interpolated to another contour of constant kinetic mixing parameter. This is due to the increase in $g_*(m_{A'})$ which needs to be compensated for by increasing the kinetic mixing. This illustrates that the predictions for COFI is very different from that of the particle dark photon scenario.

\subsection{UV-dominant freeze-in}
\label{subsec:UV-dominant}

\begin{figure}
  \centering
  \includegraphics[width=0.47\textwidth]{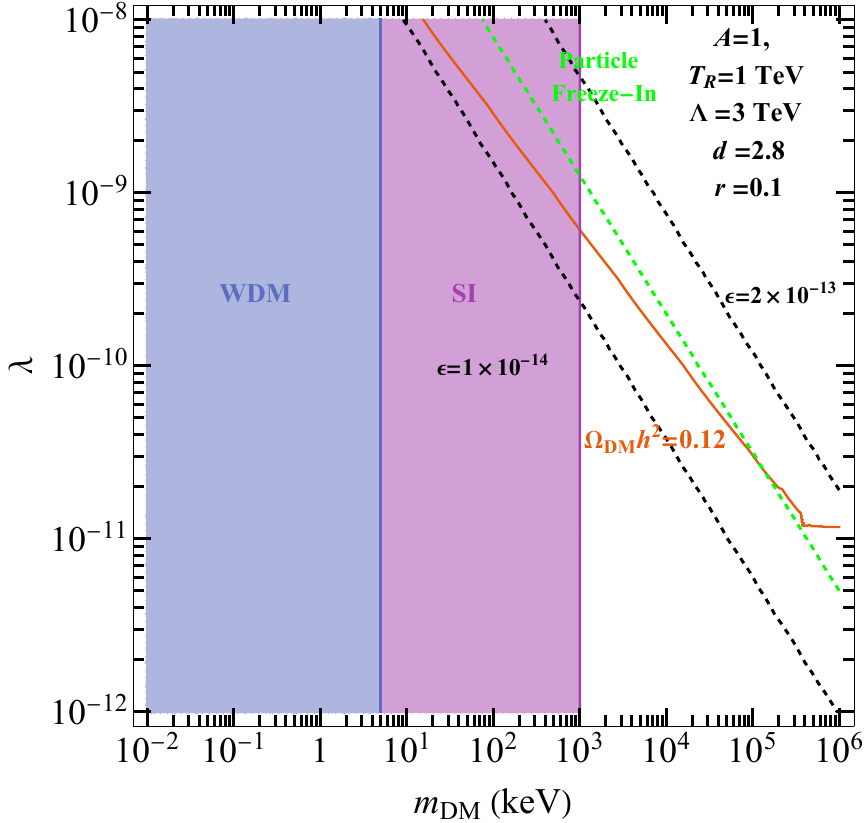}
  \includegraphics[width=0.47\textwidth]{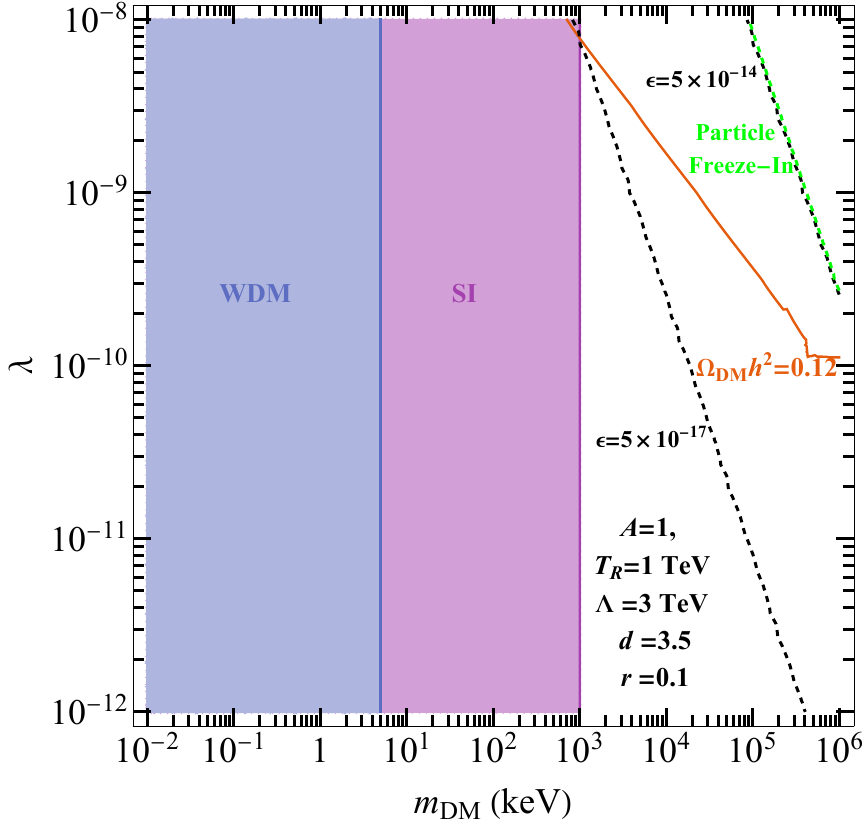}
  \caption{The dark matter mass ($x$-axis) and the SM-CFT mixing parameter, $\lambda$, ($y$-axis) which reproduces the observed dark matter relic density (orange-red solid line). The two plots have different scaling dimensions for our CFT operator; $d=2.8$ (left) and $d=3.5$ (right). For comparison, the green dashed line shows an estimate for the expected parameters which reproduces the observed relic density for the ``usual'' freeze-in with kinetic mixing parameter given by \autoref{eq:kmixing}  and the same relation between the dark photon mass and dark matter mass \cite{Blennow:2013jba}. The blue shaded region corresponds to the region of parameter space excluded by the warm dark matter bound. The purple shaded region is excluded by DM self-interaction bound coming from the observation of the bullet-cluster. The dashed curves show contours of constant kinetic-mixing parameter, $\epsilon$, with $g_s$=1.}\label{UV}
\end{figure}
The contours of the observed relic density in the plane of $( m_\text{DM}, \lambda )$ (with the remaining parameters fixed\footnote{For a fixed $d$, changes to $T_R$ and $\Lambda$ consistent with $T_R<\Lambda$ is equivalent to redefining $\lambda$.}) is shown in \autoref{UV}. In both plots, we see the same behavior: two straight lines with constant, negative slope.

The physics of this scenario is simpler to describe compared to the IR-dominant freeze-in scenario. When $d>5/2$, the $T_R$-dependent terms become the dominant contribution to the relic density at low temperatures. If the dark sector temperature increases rapidly beyond $m_\text{DM}$ by the initial energy transfer, we can safely drop the $T^{2d-5}$ term in \autoref{eq:rho_CFT_relativistic} and the energy density is of the form constant$\times T^4$. This tells us that the energy transfer from the SM sector has concluded and the dark sector energy density simply redshifts as radiation. This continues until the dark matter becomes non-relativistic, which occurs at
\begin{equation}
  T_\text{NR}=\left[\frac{\text{m}_\text{pl}}{A\sqrt{g_*(T_R)}(2d-5)}B_d(\text{total})T_R^{2d-5}\right]^{-1/4}m_\text{DM}.
\end{equation}
The energy density then continues to redshift as matter until today. This gives
\begin{equation}
  \rho_{\text{ds},0}=\left[\frac{\text{m}_\text{pl}}{A\sqrt{g_*(T_R)}(2d-5)}B_d(\text{total})T_R^{2d-5}\right]^{3/4}A m_\text{DM}T_0^3
\end{equation}
Factoring out the $m_\text{DM}$ and $\lambda$ dependence, we see
\begin{equation}
  \rho_{\text{ds},0}\propto\lambda^{3/2}m_\text{DM}.
\end{equation}
So a log-log contour plot of constant $\rho$ will look like a straight line. 

In the case where the dark sector was never relativistic, the energy density today is simply
\begin{equation}
  \rho_{\text{ds},0}=\frac{\text{m}_\text{pl}}{\sqrt{g_*(T_R)}(2d-4)}B_d(\text{total})T_R^{2d-4}T_0^3 \propto m_\text{DM}^0
\end{equation}
Here, we see that the relic density is independent of the dark matter mass\footnote{Just like in the case of the IR-dominant production, there are higher order corrections which do depend on the dark matter mass.}. Thus, a contour plot of fixed $\rho$ is simply a flat line in $\lambda$

\subsection{Phenomenological constraints}
\label{subsec:constraints}

In this section, we discuss experimental as well as theoretical constraints on our model. Ones with non-trivial restrictions on the allowed parameter space are included in the plots of our main results \autoref{Fig:relic_density_constraints_IR_dom} and \autoref{UV}.

\subsubsection{Non-equilibrium}
The dark sector must be out of equilibrium with the SM for the freeze-in assumption to be valid. Otherwise, the backreaction from the CFT to SM sector must be included in the Boltzmann equation. For this to be true, we must have
\begin{equation}
  \Gamma=n\langle \sigma v\rangle<H
\end{equation}
Using dimensional analysis, the LHS is roughly
\begin{equation}
\Gamma \sim \lambda^2\frac{T^{2d-3}}{\Lambda^{2d-4}},
\end{equation}
while the RHS is roughly 
\begin{equation}
 H \sim \frac{T^2}{\text{m}_\text{pl}}
\end{equation}
Rearranging gives
\begin{equation}
  \lambda^2\lesssim\left(\frac{\Lambda}{T}\right)^{2d-4}\frac{T}{\text{m}_\text{pl}}.	
\end{equation}
For IR-dominant production (i.e. $2 < d < 5/2$), it is sufficient to demand that it was out of equilibrium at the very last moment.
A ballpark estimation can be made by considering $m_\text{DM} \approx m_e$ and $d=2.25$, $\Lambda=1\text{ TeV}$ for which we get
$\lambda\lesssim 10^{-10}$. The above bound can be translated to a bound on the kinetic mixing

\bea
&& \epsilon \lesssim \frac{1}{g_s} \left( \frac{m_e}{\text{m}_\text{pl}} \right)^{1/5} \left( \frac{m_{\rm gap}}{m_e} \right)^{d-2} \;\;\; {\rm for} \; m_\text{DM} < m_e, \\
&& \epsilon \lesssim \frac{1}{g_s} \left( \frac{m_\text{DM}}{\text{m}_\text{pl}} \right)^{1/5} \left( \frac{m_{\rm gap}}{m_\text{DM}} \right)^{d-2} \;\;\; {\rm for} \; m_\text{DM} > m_e,
\eea
where $g_s \sim \mathcal{O} (1)$ is the coupling among hadrons of confining phase of the dark CFT.

For UV-dominate production (i.e. $d>5/2$), exponent of $T$ is positive. So we need non-equilibrium to hold at the onset. So this translates to 
\begin{equation}
  \lambda^2<\left(\frac{\Lambda}{T_R}\right)^{2d-4}\frac{T_R}{\text{m}_\text{pl}}
\end{equation}
Choosing $\Lambda\sim T_R =1\text{ TeV}$, we get $\lambda\lesssim 10^{-8}$.

\subsubsection{DM self-interaction}
\label{subsubsec:DM_SI}

Once the CFT confines, we expect that hadrons of the IR phase interact with each other with coupling strength $g_s$.  
Unlike the scenario studied in \cite{Hong:2019nwd}, we must have vector-boson-mediated self interactions in the dark sector. Here, the vector-meson is nothing but the composite dark photon coming from the CFT operator $\mathcal{O}_{\mu\nu}$. The estimate of the cross section via dimensional analysis is given by
\begin{equation}
  \sigma_\text{self}\sim \frac{1}{8\pi} \frac{m_\text{DM}^2}{m_\rho^4} = \frac{1}{8\pi m_{\rm gap}^2} r^2.
\end{equation}
In this case (i.e~dimension 4 vector mediation as opposed to dimension 5 scalar mediation), we see that the suppression is only $r^2$ rather than $r^6$ as in \cite{Hong:2019nwd}. The DM self-interaction bound 
\beq
\frac{\sigma_{\text{self}}}{m_\text{DM}} < 4500 \; {\rm GeV}^{-3}
\eeq
becomes
\beq
m_\text{DM} \gtrsim  
10 r^{4/3} \; {\rm MeV}.
\eeq

\subsubsection{Warm dark matter}

For $m_\text{DM}\sim\text{keV}$, our dark sector generically starts of as a relativistic plasma. Therefore, one needs to worry about a potentially large free streaming distance ($\lambda_\text{FS}$); suppressing structure formation below that length scale. Assuming collisionless dark matter, $\lambda_\text{FS}$ is the comoving distance traveled until some late time when the dark matter becomes highly non-relativistic. Following original derivation in \cite{Hong:2022gzo}, we know that the mean-free path in COFI theories are given by
\begin{equation}
  \lambda_\text{FS, bound}\sim \left.\frac{1}{T}\right|_{T_\text{ds} =m_\text{DM}} = \frac{1}{T_{\rm NR}}.
\end{equation}
By demanding the correct relic abundance, we can use \autoref{eq:postprod} to write $T_\text{NR}$ as a constant times $m_\text{DM}^{4/3}$. With that constraint, the warm dark matter bound is constant in the dark matter mass\footnote{While the warm dark matter bound will be shown as a constant in $m_\text{DM}$ in our plots, it is important to note that the relic density for all points above the orange-red line is less than the observed relic density. This implies that the dark sector is colder. In addition, as our dark matter is now a subcomponent, the warm dark matter bound is in-principle further relaxed. As such, the shaded region above the line should be interpreted as a conservative estimate for the exclusion.}. 

Depending on the details of the modelling, the constraints on the mass of a warm thermal relic is given by \cite{Irsic:2017ixq}
\begin{equation*}
  m_\text{DM}\gtrsim 3.5 - 5.3\text{ keV}.
\end{equation*}

\subsubsection{Constraints from searches at terrestrial experiments}
In our model, the dark photon almost always decays invisibly. The dominant constraints from dark photons to invisible searches and LDMX projections are quoted in \cite{Berlin:2018bsc}. Since the kinetic mixing required to satisfy the out-of-equilibrium constraint is sufficiently small, we safely evade these bounds.
Furthermore, as the dark photon is relatively heavy, we do not obtain a $1/v^2$ enhancement for direct detection experiments.

\subsubsection{Stellar evolution}
For a given stellar system with internal processes occurring at a scale $T_S$, there can be very distinct phenomenology depending on the mass of the dark matter candidate. If $m_\text{DM}>T_S$, we are in the hadronic phase which prevents CFT states from being directly produced. Furthermore, the dark matter particle cannot be produced directly as it is kinematically forbidden. If $m_\text{DM}\lesssim T_S \lesssim m_\text{gap}$, then the dark matter production is no longer kinematically forbidden. Lastly, if $T_S \gtrsim m_\text{gap}$, we are in the CFT phase, so CFT states are produced within the star. Only the latter two scenarios can rule out regions of parameter space. In the following subsections, we will briefly discuss the expected features of the constraints obtained via these two scenarios. The details of the estimate will be presented in \autoref{app:star_cooling} and we refer to \cite{Hong:2022gzo} for a general discussion of star cooling bounds in COFI theories.

\noindent\textbf{(i)} $T_S \gtrsim m_\text{gap}$

In this region, the only parameters of the model that influence the energy density loss rate, $\dot{\varepsilon}$, are $\Lambda,~\lambda,$ and $d$. Provided that $\Lambda$ and $\lambda$ are kept fixed, any constraints on $\dot{\varepsilon}$ will translate to a constant upper bound on $d$. An order-of-magnitude estimate on $\dot{\varepsilon}$ was performed and yielded an upper bound below the unitarity limit of $d$ for all stellar systems. As such, this feature will not be seen on the plots. However, it should be noted that $\mathcal{O}(1)$ factors in $\dot{\varepsilon}$ can have sizable impacts on the bound on $d$. This could potentially alter our conclusion on the bounds for HB stars but not the other systems.

\noindent\textbf{(ii)} $m_\text{DM}\lesssim T_S \lesssim m_\text{gap}$

In this scenario, $\dot{\varepsilon}$ depends on all of the parameters of the model. In particular, as all of the dark matter production is mediated via the dark photon, the novel energy loss rate will always have the following dependence
\begin{equation}
  \dot{\varepsilon}\propto\left(\frac{\lambda m_\text{gap}^{d-2}}{\Lambda^{d-2}}\right)^2\frac{1}{m_\text{gap}^4}
\end{equation}
As $m_\text{gap}\ll\Lambda$, $\dot{\varepsilon}$ is a decreasing function of $d$. When $2\leq d <5/2$, $\dot{\varepsilon}$ is also a decreasing function of $m_\text{gap}$. So the curve of constant $\dot{\varepsilon}$ will have negative slope on the $m_\text{gap} \text{--} d$ plane (and by extension, the $m_\text{DM} \text{--} d$ plane).

The systems which provide the strongest constraints are HB stars and RG cores. Both of these systems facilitate scattering processes with $T_S \sim 10$ keV; covering $10r$ keV$\lesssim m_\text{DM}\lesssim 10\text{ keV}$. The total allowed ``novel'' energy loss rate for these systems is typically constrained to be within the neutrino flux. Numerically, the constraints from both systems are derived from the “effective Fermi constant”. So these constraints are both comparable.

MS stars provide a much weaker constraint. The total allowed ``novel'' energy loss is orders of magnitudes larger than the solar neutrino flux. This results in a much weaker constraint at $r\text{ keV}\lesssim m_\text{DM}\lesssim 1\text{ keV}$.

SN1987A does not provide any constraint for $30r\text{ MeV}\lesssim m_\text{DM}\lesssim 30\text{ MeV}$. In the hadronic phase, our model is the usual dark photon particle freeze-in scenario with a heavy dark photon. This tells us that the novel energy loss is the same as the usual dark photon freeze-in models with an additional $(T_S /m_\text{gap})^4$ suppression. Given that SN1987A does not constrain particle freeze-in, this is also true for us.

\section{Conclusion} 
\label{sec:conclusion}

In this paper, we have considered a scenario in which a dark sector is described by a CFT and it interacts with the Standard Model via an antisymmetric tensor coupling
\beq
\mathcal{L} \supset \frac{\lambda}{\Lambda^{d-2}} B_{\mu\nu} \mathcal{O}^{\mu\nu},
\label{eq:COFI_conclusion}
\eeq
where $B_{\mu\nu}$ is the field strength of the $U(1)_Y$ gauge boson of the SM and $\mathcal{O}^{\mu\nu}$ is an antisymmetric tensor operator of the dark CFT. Provided the coupling is sufficiently small, we show that the dark sector can be populated via Conformal Freeze-In \cite{Hong:2019nwd, Hong:2022gzo}. In our case, the freeze-in production is through a tensor (as opposed to scalar) coupling. A successful implementation of the freeze-in mechanism also requires the reheating to be preferential to the SM sector.  We propose a scenario involving a cascade of CFTs, ending with the CFT describing the dark sector. This model provides a dynamical explanation of the hierarchy of scales, sizes of the couplings, as well as a natural realization of the asymmetric reheating. 

Once the dark CFT confines, a composite dark photon emerges from the above coupling with a highly suppressed kinetic mixing with the $U(1)_Y$ gauge boson. This composite dark photon couples to a dark matter particle, which we assume to be a Goldstone boson of a spontaneously broken global symmetry. The size of kinetic mixing also has a unique positive correlation with the mass gap scale; hence the mass of the dark photon \autoref{eq:kmixing}.

All these features combined make the theory very predictive and, at the same time, represent an example where small couplings and mass scales significantly different from the ones appearing in the SM are explained rather than just assumed as inputs.

We study in detail the dark matter production, cosmological evolution, and relevant constraints from considerations of dark matter self-interaction, warm dark matter bound, and stellar evolution. We consider both possibilities where the dark matter production is UV- and IR-dominant, and show that the correct relic abundance can be obtained with reasonable choices of parameters. We found viable dark matter candidates in the range of MeV to GeV, with a dark sector confinement scale and dark photon mass  a factor of approximately 10 --100 times higher.  

There is one important distinction between our setup and the ``usual" scenario with only an elementary dark photon with a tiny coupling to the SM.  In our setup, the relic abundance is mainly determined by the conformal dynamics instead of being mediated by the dark photon. Hence, it leads to very different predictions for the correlation between the dark matter and dark photon properties. A richer dark sector can lead to richer physics as well. For example, the cascade of phase transitions between the transitions among the CFTs can leave their imprints in cosmological observations, such as the  gravitational wave  signals. We leave further exploration of these interesting possibilities for a future study. 

\section*{Acknowledgments}

We are very grateful to Nima Afkhami-Jeddi, Kaustubh Agashe, Jae Hyeok Chang, Seth Koren, and Julio Virrueta for helpful discussions and Seth Koren for collaboration in the early stage of the project.
SH thanks Gowri Kurup and Maxim Perelstein for useful discussions and collaboration on a related  subject.
This work was initiated and performed in part at Aspen Center for Physics, which is supported by National Science Foundation grant PHY-1607611.
WHC acknowledges the University of Chicago Francis and Rose Yuen Campus for its hospitality during the final phase of this study.
S.H.\ has been partially  supported by the U.S.~Department of Energy under contracts  No.DE-AC02-06CH11357 at Argonne National Laboratory. The work of LTW, WHC, and the work of S.H. at the University of Chicago has been supported by the DOE grant DE-SC-0013642. S.H. and LTW would like to thank Aspen Center for Physics, where this work is initiated, for hospitality.

\appendix

\section{Dynamical Mass Scale Generation in COFI Theories}
\label{sec:dsScales}

\subsection{Gap scale in COFI theories with a scalar operator}
\label{subsec:gap_scalar_COFI}

In the UV, COFI theories assume a coupling between the SM sector and a CFT sector of the form
\beq
\mathcal{L} \supset \frac{\lambda}{\Lambda^{D-4}} \mathcal{O}_{\rm SM} \mathcal{O}_{\rm CFT}
\label{eq:COFI_setup}
\eeq
where $\mathcal{O}_{\rm SM/CFT}$ is a gauge invariant operator (invariant under its own gauge group) and $D$ is the sum of scaling dimensions of the two operators. The dimensionless coupling $\lambda$ needs to be small for the dark matter relic density to be obtained via freeze-in. Such a small value can arise naturally from dimensional transmutation if the above theory emerges as an IR phase of a UV gauge theory with a IR-fixed point. See \cite{Hong:2019nwd} for a UV completion via weakly coupled gauge theory and \autoref{sec:setup} for a UV completion in terms of strongly coupled CFT with a 5d holographic dual. 

In the absence of other conformal symmetry breaking terms, the interaction in \autoref{eq:COFI_setup} is the main source of conformal symmetry breaking. The details of how this occurs depend on the nature of the SM operator $\mathcal{O}_{\rm SM}$. If the vacuum expectation value (vev) of $\mathcal{O}_{\rm SM}$ is non-zero, then the renormalization group (RG) flow ensures that the above theory flows to 
\beq
\mathcal{L} \sim \frac{\lambda}{\Lambda^{D-4}} \langle \mathcal{O}_{\rm SM} \rangle \mathcal{O}_{\rm CFT}
\eeq
at $E \sim \langle \mathcal{O}_{\rm SM} \rangle$.
This can be recognized as a scalar deformation to the CFT and triggers running of the CFT (provided the CFT operator has dimension $\leq 4$). The scale at which the conformal invariance is completely lost and a new IR phase (we assume that it is the usual confinement phase) arises is estimated to be \cite{Hong:2019nwd}
\beq
m_{\rm gap} \sim \left( \frac{\lambda}{\Lambda^{D-4}} \langle \mathcal{O}_{\rm SM} \rangle \right)^{1/(4-d)},
\eeq
where $d$ is the scaling dimension of the CFT operator. Below the gap scale (i.e. $E \leq m_{\rm gap}$), CFT states turn into composite particles states. Some of these composite states may be stable on a cosmological time scale and plays the role of DM. 

Even when $\langle \mathcal{O}_{\rm SM} \rangle = 0$, conformality loss still occur due to ``operator-mixing effects''\cite{Hong:2022gzo}. The idea is that given the coupling in \autoref{eq:COFI_setup}, other sets of interactions are induced either at tree or loop level. This gives
\beq
\mathcal{L} \sim \sum_i b_i \mathcal{O}_{\rm CFT} + c_i \mathcal{O}^i_{\rm SM} \mathcal{O}_{\rm CFT} + d_i \mathcal{O}_{\rm CFT}^2,
\label{eq:operator_mixing}
\eeq
where $b_i$, $c_i$ and $d_i$ are generic dimensionful coefficients which can be reliably estimated within a theory. 

\begin{figure}[h]
  \centering
  \includegraphics{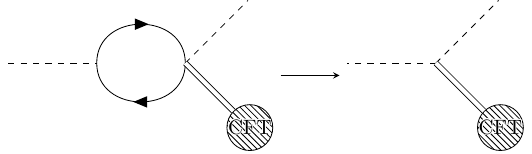}
  \caption{Example of the operator mixing effect}
  \label{fig:op_mixing}
\end{figure}

The first kind of mixing effect with a coefficient $b_i$ arises by contracting all SM fields in $\mathcal{O}_{\rm SM}$ forming a loop diagram. An important example is the gluon portal with $\mathcal{O}_{\rm SM} = G_{\mu\nu}^a G^{a \mu\nu}$. In this case, one simply closes up gluon lines in a loop and it provides a dominant source for the CFT-breaking \cite{Hong:2022gzo}. For the second kind with a coefficient $c_i$, some of the induced operators $\mathcal{O}_{\rm SM}^i$ can have non-vanishing vev leading to the breaking of conformal symmetry as described in the procedure above. For instance, starting with $\mathcal{O}_{\rm SM} = HL^\dagger  \ell_R$, the operator $ (H^\dagger H) \mathcal{O}_{\rm CFT}$ is generated at one-loop as shown in \autoref{fig:op_mixing}. This provides a source of CFT breaking. In addition, $\mathcal{O}_{\rm CFT}^2$ may contribute to the breaking of conformal invariance if its scaling dimension is not greater than $4$ (in the large-$N$ limit, the scaling dimension of $\mathcal{O}_{\rm CFT}^2$ is roughly twice that of $\mathcal{O}_{\rm CFT}$). Ultimately, the mass gap scale is obtained by taking into account of all these effects, and is primarily determined by the largest breaking effect. See \cite{Hong:2022gzo} for a comprehensive discussion.

\subsection{Gap scale in COFI theory with an anti-symmetric tensor operator}
\label{subsec:gap_tensor_COFI}

At a UV scale, $\Lambda$, on the order of a few TeV, the Lagrangian of the theory is given by
\beq
\mathcal{L} \sim \frac{\lambda}{\Lambda^{d-2}} B_{\mu\nu} \mathcal{O}^{\mu\nu}
\label{eq:COFI_Dark_Photon}
\eeq
where $B_{\mu\nu}$ is the field strength of $U(1)_Y$ gauge boson of the SM. 
For freeze-in production of DM, we take $\lambda \sim \mathcal{O} (10^{-10} {\text{--}} 10^{-11})$. This small coupling can arise naturally according to the construction described in \autoref{sec:setup}.

In order to study the RG evolution of the theory described by \autoref{eq:COFI_Dark_Photon} and the gap scale generation, we first note that no operator mixing effect of the first two kinds ($b_i$ and $c_i$ terms in \autoref{eq:operator_mixing}) can lead to a reliable source for the conformal symmetry breaking. This is simply because such induced operators are not scalar CFT operators. 

Moving onto the third kind, at scales above the vev of the SM Higgs, $v$, the operator in \autoref{eq:COFI_Dark_Photon} does not mix with any other \emph{local scalar} operators proportional to $\mathcal{O}_{\mu\nu} \mathcal{O}^{\mu\nu}$. This is because any diagram superficially generating such an operator involves massless propagator of the hypercharge gauge boson $B_\mu$ and hence is non-local. This changes once electroweak symmetry is broken. Now, $B_\mu = \cos \theta_W A_\mu + \sin \theta_W Z_\mu$. At $E < m_Z$, the exchange by $Z$-boson can generate other local operators.
To see this we consider a scalar operator from OPE of two $\mathcal{O}_{\mu\nu}$'s
\beq
\mathcal{O}_{\mu\nu} \times \mathcal{O}^{\mu\nu} \supset \left[ \mathcal{O}_{\mu\nu} \mathcal{O}^{\mu\nu} \right]_{n,\ell} \sim \mathcal{O_{\mu\nu}} \Box^n \partial_{\mu_1} \cdots \partial_{\mu_\ell} \mathcal{O}^{\mu\nu}
\eeq
and the scaling dimension of the scalar operator $\mathcal{O}_s$ in the OPE expansion (i.e.~$n=0, \ell=0$) is given by $d_s = 2 d + \gamma$ where $\gamma$ is the anomalous dimension. To the extent that negative anomalous dimension is possible, it is expected that such an operator can serve as a scalar deformation to the CFT.\footnote{In AdS/CFT, the anomalous dimension $\gamma_{n,\ell}$ of a general spinned operator in the OPE expansion corresponds to the binding energy of the two antisymmetric tensor particles in the bulk. The scaling dimension is dual to the bulk energy of such a bound state and is given by $\Delta = 2 d + 2n + \ell + \gamma_{n, \ell}$, where $\Delta$ and $\ell$ are the scaling dimension and the spin of the operator in the OPE expansion respectively. In the large-$\ell$ limit, it is known that the anomalous dimension takes the universal behavior $\gamma_{n,\ell} \sim \ell^{-\tau}$ where the twist $\tau$ is defined by $\tau = \Delta - \ell$. In particular, the energy momentum tensor (which exists in any QFT) satisfies $\tau = 2$ and $\gamma_{0,2} < 0$; the latter being the dual of the fact that the gravitational force is attractive. See e.g.~\cite{Jared:lecture} for more discussion. } \\
In fact, the $Z$-boson exchange generates the operator
\beq
\mathcal{L} \sim \left( \frac{\lambda}{\Lambda^{d-2}} \right)^2 \frac{e_s \sin^2 \theta_W}{m_Z^{d_s-2d}} \mathcal{O}_s
\label{eq:Os_OPE_1}
\eeq
where the scalar operator $\mathcal{O}_s$ is the lowest dimensional operator with dimension $d_s$ in the OPE with dimensionless coefficient $e_s$.

In this work, we assume that there exists a CFT scalar operator $\mathcal{O}_s$ with scaling dimension $d_s$ and there is a large gap in the CFT operator spectrum such that the scale of conformality lost is reliably estimated by the RG running of this single operator. To the best of our knowledge, no numerical CFT bootstrap bound on the scaling dimension of such a scalar operator from the OPE of antisymmetric rank-2 tensor operator is available in the literature. It would be interesting to compute the bound and to see if non-trivial constraints on our scenario is imposed.

Given a scalar deformation term
\beq
\mathcal{L} \sim c_s \mathcal{O}_s,
\label{eq:CFT_deformation}
\eeq
the gap scale is estimated to be $m_{\rm gap} \sim c_s^{1/(4-d_s)}$. Since we do not have prior knowledge of the scalar deformation generated from the OPE, we simply treat $m_{\rm gap}$ as a free parameter for our study of the dark matter phenomenology.

\section{AdS/CFT Correspondence for COFI}
\label{app:AdS/CFT}

\subsection{Details of the 5d dual}

In this section, we discuss the AdS dual of the theory setup described in \autoref{sec:setup}.

\begin{figure}[h]
  \centering
  \includegraphics[width=0.75\textwidth]{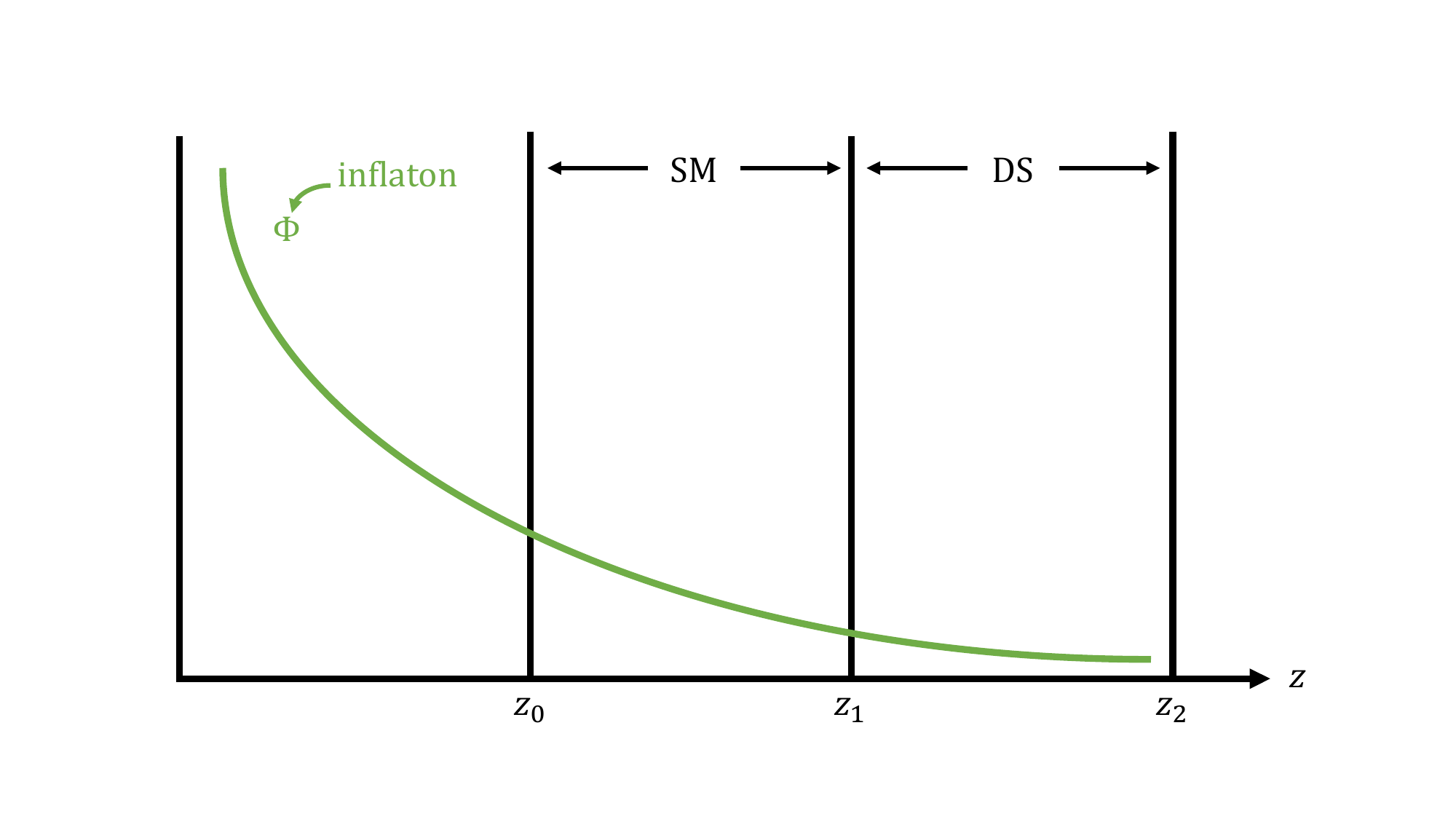}
  \caption{$\text{AdS}_5$ picture corresponding to the 4d COFI theory setup in \autoref{sec:setup}.}
  \label{Fig:AdS-CFT1}
\end{figure}

A 4d theory of COFI can be thought of as the dual of a theory living on a slice of $\text{AdS}_5$. A simple cartoon level of this $\text{AdS}_5$ picture is depicted in \autoref{Fig:AdS-CFT1}. Neglecting the first part of the bulk associated with the physics of inflation, roughly speaking, there is a bulk where all the SM fields propagate as in the standard Randall-Sundrum (RS) model \cite{Randall:1999ee, Randall:1999vf}, which is dual to a CHM in 4d. There exists an additional bulk in the deeper IR (i.e.~larger $z$) where dark sector (DS) fields propagate.\footnote{The theoretical framework for this type of generalization of the standard RS model with multiple branes was introduced in \cite{Agashe:2016rle} and phenomenology was studied in \cite{Agashe:2016kfr, Agashe:2017wss, Agashe:2018leo, Agashe:2020wph}.} The two sectors communicate via brane-localized interactions.\footnote{In COFI, for simplicity, the SM sector is taken to be purely elementary. This may be realized by taking a limit in 5d in which the SM-bulk is taken to be a infinitely thin brane.}  

\begin{figure}
  \centering
  \includegraphics[width=0.75\textwidth]{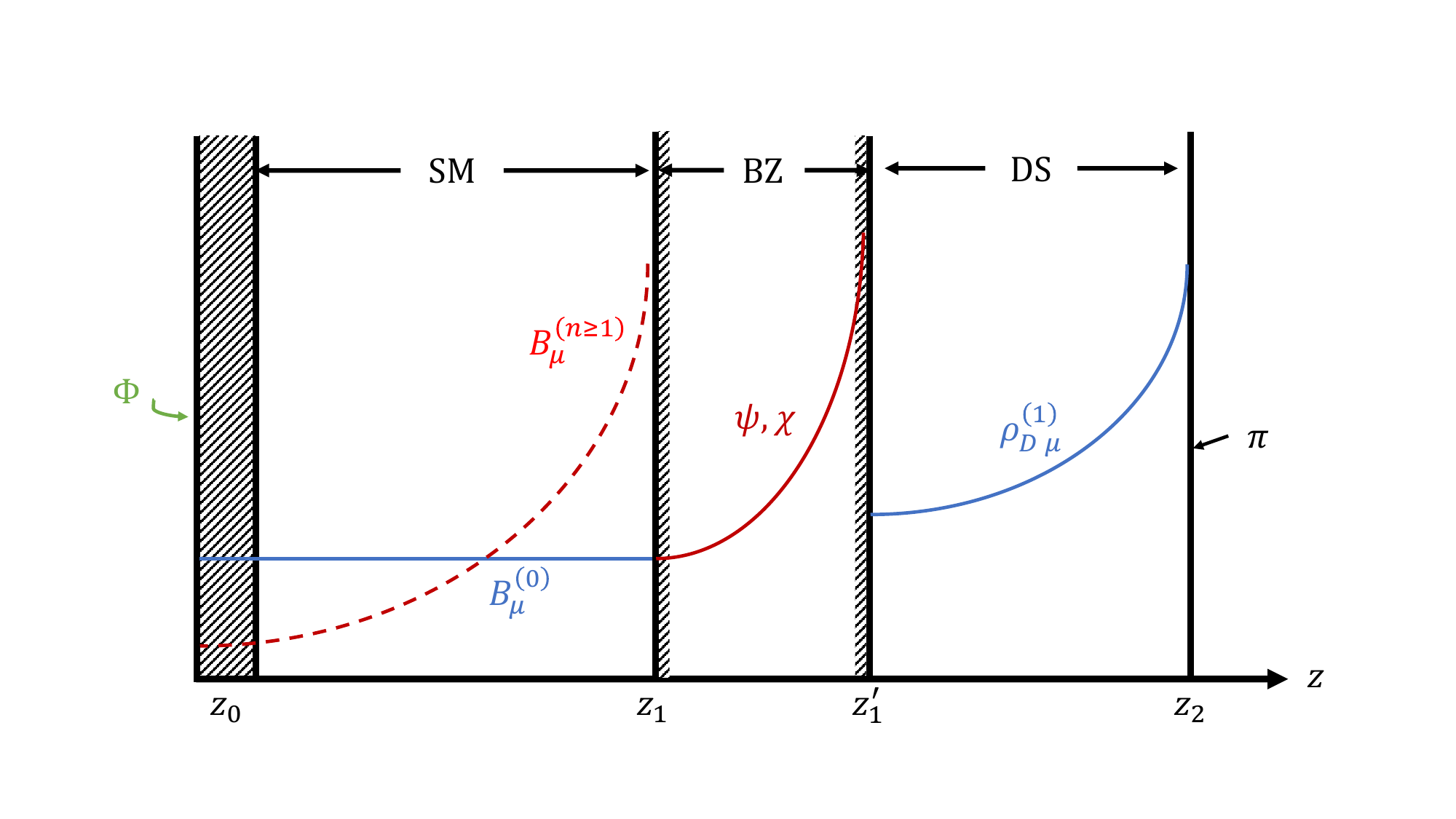}
  \caption{$\text{AdS}_5$ picture leading to a small portal coupling $\lambda$ appearing in 4d CFT picture. An extra bulk (called BZ-bulk) is introduced between the SM and DS bulk. The interactions between the BZ-bulk fields with SM and DS states generate desired suppression in the effective coupling between the SM and DS. The inflaton sector is simplified to a thin-brane picture. $\pi$ represents the DM state(s).}
  \label{Fig:AdS-CFT2}
\end{figure}

Inflation occurs at a very high energy scale, and so it is natural that the inflaton appears in the most UV part (small $z$) of the theory in 5d.
In \autoref{Fig:AdS-CFT1}, we added a ``sector'' of inflation depicted as an extra bulk slice beyond the SM slice. If the profile of the inflaton field, $\Phi$, is inclined towards the UV brane, a completely natural picture emerges in which the SM sector gets reheated much more than the dark sector simply by the size of overlap with the inflaton field; in 5d, this is a consequence of the geometric (de)localization and in 4d, it is dual to the renormalization group flow effects. In order to simplify the discussion, we note that the details within the ``inflation-bulk'' is not important for us and we will simply take a thin brane limit for the inflaton sector (see \autoref{Fig:AdS-CFT2} but we still use \autoref{Fig:AdS-CFT1} for the discussion below). 

The presence of a throat further in the deep IR (i.e. beyond $z=z_1$) in the 5d dual means that in 4d the confinement of $\text{CFT}_1$ at $\Lambda_1$ (associated with $z_1$) also creates a set of interacting \emph{composite preons} (like quarks and gluons of the QCD). This sector carries no SM charges and their dynamics brings the sector into a strongly interacting IR fixed point at a scale not so much below $\Lambda_1$. This course of physics is not quite spelled out in the 5d physics when represented as a thin brane separating the two bulks. 

In 5d, we add a $U(1)_{\scriptscriptstyle \rm D}$ gauge field in the DS-bulk and choose $(+,-)$ boundary conditions (BCs) (i.e.~Neumann BC on the intermediate brane and Dirichlet BC on the IR brane). This ensures that there are no zero modes and at the same time allows us to write down a brane-localized interaction. The brane-localized interactions takes the form
\beq
\mathcal{L}_{\rm brane} \sim \epsilon_{\scriptscriptstyle \rm 5d} B_{\mu\nu} F_{\scriptscriptstyle \rm D}^{\mu\nu} = \epsilon_{\scriptscriptstyle \rm 5d} \sum_n \sum_m f_B^{(n)} (z_1) f_{\scriptscriptstyle \rm D}^{(m)} (z_1) B_{\mu\nu}^{(n)} F_{\scriptscriptstyle \rm D}^{(m) \mu\nu}.
\label{eq:COFI_5d}
\eeq

Since the $U(1)_Y$ KK-modes, $B_{\mu}^{(n>0)}$, have profiles localized to the intermediate brane at $z=z_1$, they have sizable coupling with the dark $U(1)_{\scriptscriptstyle \rm D}$ gauge boson $A_{\scriptscriptstyle \rm D}^{(1)}$.\footnote{To be more precise, the profile of $A_{\scriptscriptstyle \rm D}^{(1)}$ is peaked near the IR brane and suppressed at the intermediate brane. This may raise a question of whether the effective coupling is highly suppressed. From the discussion of elementary-composite mixing given in \autoref{subsec:UV theory}, however, we know that this suppression is only $\mathcal{O} (g/g_s)$.} If $T_R \sim \Lambda_1$ (but less than confinement phase transition temperature), these KK modes can be excited and can easily populate the dark sector. If $T_R \ll \Lambda_1$, they are not produced cosmologically; leaving only the zero mode, $B_\mu^{(0)}$, (i.e.~SM field) coupled to the dark CFT. If, on the other hand, $T_R > \Lambda_1$, a more appropriate description is in terms of the thermal $\text{CFT}_1$ in which there is no clear distinction between the SM and the DS. 

Next, we show that small $\lambda$ appearing in \autoref{eq:Theory_COFI} requires the existence of an extra bulk (BZ-bulk) between the SM and DS bulk as shown in \autoref{Fig:AdS-CFT2}. 
Let us first discuss the case with only SM and DS bulks (i.e.~the BZ-bulk is shrunk to a thin brane). The $U(1)_{\scriptscriptstyle \rm D}$ gauge field living in the DS-bulk couples to the SM sector by a kinetic mixing written down on a brane, $B_{\mu\nu} F_{\scriptscriptstyle \rm D}^{\mu\nu}$. In the CFT picture, this means that the first confinement at $\Lambda_1$ gives rise to a ``composite'' CFT (called it $\text{CFT}_{\scriptscriptstyle \rm D}$ in \autoref{sec:setup}) and it interacts with the SM fields via
\beq
\mathcal{L} \sim \mathcal{L}_{\rm CFT_{\scriptscriptstyle \rm D}} + \epsilon \rho_{B \mu\nu} F_{\scriptscriptstyle \rm D}^{\mu\nu} + g_{\scriptscriptstyle \rm D} A_{{\scriptscriptstyle \rm D} \mu} J^\mu_{\scriptscriptstyle \rm D} 
\label{eq:app_AdSCFT_Lambda1}
\eeq
where $\rho_B^\mu$ is a composite vector meson associated with $U(1)_Y$ and $A_{\scriptscriptstyle \rm D}^\mu$ and $F_{\scriptscriptstyle \rm D}^{\mu\nu}$ is a composite vector boson and its field strength external to the dark CFT. The latter couples to the dark CFT through its coupling to $U(1)_{\scriptscriptstyle \rm D}$ current and to the composite SM sector via kinetic mixing with $\rho_B^\mu$. 
Since $\text{CFT}_{\scriptscriptstyle \rm D}$ is purely composite, its interaction with the external field $B_\mu$ needs to be through its coupling to a composite state, $\rho_B^\mu$, which then mixes with $B_\mu$. The composite-elementary mixing is $\mathcal{O} (g/g_{1s})$, where $g (g_{1s})$ is the elementary (composite) gauge coupling\footnote{In more detail, the composite-elementary mixing is of the form $\frac{g}{g_{1s}} \Lambda_1^2 B_\mu \rho_B^\mu$, which originates from $g B_\mu J^\mu$ in the UV Lagrangian using the interpolation relation $J^\mu \sim \frac{\Lambda_1^2}{g_{1s}} \rho_B^\mu$. See section~2.3.2 of \cite{Agashe:2016rle} for more details. }. If the reheat temperature is less than $\Lambda_1$, \autoref{eq:app_AdSCFT_Lambda1} is the right description after the reheating. This, however, comes with a sizable interaction between the SM and DS. The SM interacts with the DS through $B^\mu \text{--} \rho_{B}^\mu$ mixing and then $\rho_B^\mu \text{--} A_{\scriptscriptstyle \rm D}^\mu$ mixing, and finally $A_{\scriptscriptstyle \rm D}^\mu$ coupling to the dark CFT. Generically, we expect that $\epsilon$ and $g_{\scriptscriptstyle \rm D}$ are not small, leading to a significant coupling which can be estimated to be
\beq
\mathcal{L} \sim \left( \frac{g}{g_{1s}} \right) \frac{ \epsilon g_{\scriptscriptstyle \rm D}}{\Lambda_1^{d-2}} B_{\mu\nu} \mathcal{O}_{\scriptscriptstyle \rm D}^{\mu\nu},
\eeq
where $d$ is the scaling dimension of the tensor operator $\mathcal{O}_{\scriptscriptstyle \rm D}^{\mu\nu}$ of the $\text{CFT}_{\scriptscriptstyle \rm D}$. We assumed that there is a coupling between $F_{\scriptscriptstyle \rm D}^{\mu\nu}$ and $\mathcal{O}_{\scriptscriptstyle \rm D}^{\mu\nu}$ which is on the order of $g_{\scriptscriptstyle \rm D}$. We see that other than the mild suppression factor from the composite-elementary mixing, the net interaction is unsuppressed and the DS will be quickly equilibrated with the SM, invalidating both the asymmetric reheating and non-thermal freeze-in production. 

To resolve this, we now introduce an extra bulk, the BZ-bulk, between the SM and DS bulks, as depicted in \autoref{Fig:AdS-CFT2}. Intuitively, this BZ-bulk can be alternatively thought as a thick opaque brane. Due to the finite penetration depth, both $B_\mu$ and $A_{{\scriptscriptstyle \rm D} \mu}$ are attenuated, resulting in an extra reduction in their overlap. More explicitly, the interaction between the SM and DS is mediated by a field living in the BZ-bulk. For instance, it can be a pair of bulk fermions, $\psi$ and $\chi$, coupling to each side via dipole interactions
\beq
S \supset \int_{z=z_1} \frac{a}{\Lambda_1} B_{\mu\nu} \bar{\psi}_L \sigma^{\mu\nu} \chi_R (z_1) + \int_{z=z_1'} \frac{b}{\Lambda_2} F_{{\scriptscriptstyle \rm D} \mu\nu} \bar{\chi}_L \sigma^{\mu\nu} \psi_R (z_1'),
\label{eq:AdSCFT_5d_BZ_bulk_action}
\eeq
where $a$ and $b$ are dimensionless constants. In order to get the above interactions, we have chosen the following boundary conditions for the bulk fermions.
\beq
\psi = \left( 
\begin{array}{cc}
\psi_L (+,-) \\ \psi_R (-,+)
\end{array}
\right) , \;\;\;\;\;\;\;
\chi = \left( 
\begin{array}{cc}
\chi_L (-,+) \\ \chi_R (+,-)
\end{array}
\right).
\eeq
Here, $+ (-)$ denotes the Neumann (Dirichlet) boundary condition, and the above choice ensures that there are no fermion zero modes, thereby removing potential inconsistency with cosmological observations (e.g.~$\Delta N_{\rm \scriptscriptstyle eff}$).

Crucially, if the reheat temperature is below $\Lambda_2$ (dual to $z=z_1'$), one can use KK-decomposition (as opposed to thermal CFT) to show that the exponentially suppressed profile leads to a very small the effective coupling between the SM and DS. This suppression is a 5d dual version of suppression seen in 4d picture from RG running (i.e. the discussion around \autoref{eq:Theory_at_Lambda_2} and \autoref{eq:effective_mixing} and analogous discussion for $d_{L,R}<5/2$). More explicitly, if we choose bulk masses for $\psi$ and $\chi$ such that their zero mode profiles are localized near the brane at $z=z_1$ (corresponding to $d_{L,R} > 5/2$ in \autoref{subsec:UV theory}), then while $a$ can be $\mathcal{O} (1)$, due to profile suppressions, $b$ is exponentially suppressed. The effective coupling, denoted as $\tilde{\epsilon}$ in \autoref{subsec:UV theory} is proportional to the product $ab$ and hence highly suppressed. A similar argument applies to the opposite case $d_{L,R} < 2/5$: this time $b$ is $\mathcal{O} (1)$ but $a$ is exponentially suppressed. Contributions from KK modes are also suppressed because KK profiles are all very inclined towards the IR brane at $z=z_1'$. 
The 4d dual picture is discussed in \autoref{subsec:UV theory} and the diagram \autoref{Fig:CFT_diagram_effective_mixing} represents the sum of both zero- and first KK-modes (in a sense of 2-site truncation of \cite{Contino:2006nn}).

An effective coupling between the SM and DS is obtained by computing the fermion loop stretched between the $z_1$ and $z_1'$ branes which is UV-finite. The result should be on the order of what is shown in \autoref{eq:effective_mixing}.

\subsection{Summary of 5d picture}
The holographic dual picture of conformal freeze-in physics is shown in \autoref{Fig:AdS-CFT1}. The feature that dark sector bulk (denoted as DS) appears at larger $z$ (i.e.~deeper IR) compared to the SM-bulk is a reflection of the SM being external to the dark CFT sector in the 4d picture. Furthermore, the fact that inflation occurs at very high energy scale makes it natural that the ``inflation-bulk'' appears in the deepest UV (i.e.~smallest $z$). Due to the smaller overlap with the DS states, the SM states can be preferentially produced at reheating. To ensure asymmetric reheating, we need the coupling between the two sectors to be small. This necessitates another bulk (shown as ``BZ'' in \autoref{Fig:AdS-CFT2}); providing effective sequestering of the DS bulk.

\section{Hadronic Production}
\label{sec:HadProduction}

Provided $m_\text{DM} \ll m_{\rm gap}$ (in practice an $\mathcal{O}(1)$ separation suffices), at $T < m_{\rm gap}$, most of the dark hadrons decouple and the effective theory is described by
\beq
\mathcal{L} \sim \epsilon B_{\mu\nu} \rho^{\mu\nu} + g_s J_\text{DM}^\mu \rho_\mu  + e J_\mu A^\mu,   
\eeq
where $J^\mu = \bar{\psi} \gamma^\mu \psi$ is the SM fermion (e.g.~electron) current coupled to the photon, $A_\mu$, and $J_\text{DM}^\mu = \left( \pi^\dagger \partial_\mu \pi + \text{h.c.} \right)$ is the DM current coupled to the dark photon, $\rho_\mu$. As usual, the kinetic mixing can be diagonalized to get
\beq
\mathcal{L} \sim e \epsilon J^\mu \rho_\mu + g_s J_\text{DM}^\mu \rho_\mu + e J^\mu A_\mu,
\eeq
where the first term represents the coupling of the dark photon to the SM fermion current.
Since the mass of the composite dark photon is $m_\rho \approx m_{\rm gap}$, at $T < m_{\rm gap}$, we can further integrate out the dark photon and acquire a higher-dimensional operator describing the interaction between the DM and the SM
\begin{equation}
  \mathcal{L} \sim e \epsilon g_s \frac{J_\mu J_\text{DM}^\mu}{m_{\rm gap}^2}.
\end{equation}
The higher-dimensional nature of the operator reveals that the process is UV-dominant. More explicitly, the energy transfer rate from the fermion annihilation is estimated to be
\beq
\Gamma \left( \bar{\psi} \psi \to \pi \pi \right) \sim \frac{1}{8\pi} \frac{e^2 \epsilon^2 g_s^2}{m_{\rm gap}^4} T^9.
\label{eq:Gamma_Had}
\eeq 
Here, $T^6$ is from $n_\psi^2$, a factor of $T^2$ from the derivative of $\pi$, and one factor of $T$ from $E_{\rm transfer}$ (since we are computing a rate for the energy transfer).

We now show that this production is sub-dominant and therefore, to a good approximation, we can say that COFI production ends around $T \sim m_{\rm gap}$. In order to show the subdominance condition, we first consider the case where, in the UV, there was relativistic COFI production (i.e.~$T_\text{ds} > m_\text{DM}$) and $T_{\rm NR} < m_{\rm gap}$. In this case, the Boltzmann equation \autoref{eq:BE_R} can be solved using \autoref{eq:Gamma_Had} giving
\beq
\rho_{\rm ds, Had} (T< m_{\rm gap}) \approx \frac{\hat{B} \text{m}_\text{pl}}{3 \sqrt{g_*}} T^4 m_{\rm gap}^3, \;\;\; \hat{B} = \frac{e^2 \epsilon^2 g_s^2}{m_{\rm gap}^4}=\frac{e^2\lambda^2}{m_\text{gap}^4}\left(\frac{m_\text{gap}}{\Lambda}\right)^{2d-4},
\eeq
where we have kept only the leading term (valid at $T \ll m_{\rm gap}$). On the other hand, the energy density from COFI $T > m_{\rm gap}$ is
\beq
\rho_{\rm ds, R} (T < m_{\rm gap}) \approx \frac{B_d \text{m}_\text{pl}}{(5-2d) \sqrt{g_*}} T^{2d-1}.
\eeq
The ratio of the two is
\beq
\frac{\rho_{\rm ds, Had}}{\rho_{\rm ds, R}} (T < m_{\rm gap}) = \frac{e^2}{\Gamma_d} \left( \frac{T}{m_{\rm gap}} \right)^{5-2d},
\eeq
where $\Gamma_d$ was previously defined in \autoref{eq:Collision term general form}. Since $2< d < 5/2$ for IR-dominant COFI production, the above ratio is much smaller than 1. Therefore, we see that the hadronic production is insignificant when $T_{\rm NR} < m_{\rm gap}$.

Now, consider the complementary case with $T_{\rm NR} > m_{\rm gap}$. In this case, at $T > T_{\rm NR}$, the production is via the relativistic COFI process, and at $m_{\rm gap} < T < T_{\rm NR}$, it is through the process discussed in \autoref{subsubsec:COFI_NR}. Finally, at $T < m_{\rm gap}$, further hadronic production occurs. The energy density from the hadronic production is obtained by solving \autoref{eq:BE_NR} with \autoref{eq:Gamma_Had} and we found
\beq
\rho_{\rm ds, Had} (T < m_{\rm gap}) \approx \frac{\hat{B} \text{m}_\text{pl}}{4 \sqrt{g_*}} T^3 m_{\rm gap}^4.
\eeq
Contribution from the earlier production is found using \autoref{eq:rho_CFT_relativistic} and \autoref{eq:rho_CFT_NR}:
\beq
\rho_{\rm ds, COFI} (T < m_{\rm gap}) \approx \frac{B_d \text{m}_\text{pl}}{ \sqrt{g_*}} \left( \frac{1}{(2d-4)}  + \frac{1}{(5-2d)} \right) T^3 T_{\rm NR}^{2d-4} .
\eeq
The first term is from the non-relativistic COFI production ($m_{\rm gap} < T < T_{\rm NR}$) while the second term represents the relativistic COFI production ($T > T_{\rm NR}$).
The ratio is found to be
\beq
\frac{\rho_{\rm ds, Had}}{\rho_{\rm ds, COFI}} = \frac{e^2}{\Gamma_d} \frac{(2d-4)(5-2d)}{4} \left( \frac{m_{\rm gap}}{T_{\rm NR}} \right)^{2d-4}.
\eeq
Since unitarity demands $d>2$, the above ratio must be much smaller than one. Therefore, we conclude that the hadronic production makes only a small contribution to the DM energy density and hence may be ignored. This also means that while the naive kinematics suggest that the production must end at $T < m_\text{DM}$ (when it is not terminated already by SM fermion masses), in effect it ends around $m_{\rm gap}$.

\section{Conformal Freeze-In Calculations}
\label{app:COFI_computation}

In this appendix, we present details of COFI computations used in the main text.

\subsection{Fermion pair annihilation}
\label{subapp:fermion_pair_ann}

We begin by writing down the $f\bar{f}\rightarrow\text{CFT}$ matrix element
\begin{equation}
  \mathcal{M}=-\frac{2e\lambda}{\Lambda^{d-2}}\frac{1}{p^2}\bar{u}(p_2)\gamma^\mu u(p_1)p^\rho\la p|\mathcal{O}_{\mu\rho}|0\ra,
\end{equation}
where $\mathcal{O}_{\mu\rho}$ is the antisymmetric 2-tensor operator corresponding to the CFT out state with momentum $p=p_1+p_2$. Squaring and performing the spin sum in the massless fermion limit gives
\begin{equation}
  \sum|\mathcal{M}|^2=\frac{4e^2\lambda^2}{\Lambda^{2d-4}}\frac{4}{p^4}p_{1\beta}p_{2\gamma}(g^{\beta\nu}g^{\gamma\mu}-g^{\beta\gamma}g^{\mu\nu}+g^{\beta\mu}g^{\gamma\nu})p^\rho p^\alpha \la 0|\mathcal{O}^\dagger_{\alpha\nu}|p\ra\la p|\mathcal{O}_{\mu\rho}|0\ra
\end{equation}
The collision term (rate of energy transfer through scattering) is given by
\begin{equation}
  \begin{aligned}
  n_1 n_2\la\sigma_{1+2\rightarrow\text{CFT}}v_\text{rel}E_{\text{tot}}\ra=&\left(\int d\Pi_1 f(p_1)\right)\left(\int d\Pi_2 f(p_2)\right)\\
 & \hspace{-1cm} \int\frac{d^4 p}{(2\pi)^4}\rho(p^2)(2\pi)^4\delta^4(p_1+p_2-p)
  \left(\sum|\mathcal{M}|^2\right) (E_1+E_2),
  \end{aligned}
\end{equation}
where $f$ is the phase space distribution function of the incoming fermion and $\rho(p^2)$ is understood to be the appropriate normalization for the CFT state $|p\ra$. Now notice that
\begin{equation*}
\begin{aligned}
  \la\mathcal{O}^\dagger_{\mu\nu}(x)\mathcal{O}_{\rho\sigma}(0)\ra=&\int\frac{d^4p}{(2\pi)^4}\rho(p^2)\la 0| e^{-iP\cdot x}\mathcal{O}^\dagger_{\mu\nu}(0)e^{iP\cdot x}|p\ra\la p|O_{\rho\sigma}(0)|0\ra\\
  =&\int\frac{d^4p}{(2\pi)^4}\rho(p^2)\la 0|\mathcal{O}^\dagger_{\mu\nu}(0)|p\ra\la p|O_{\rho\sigma}(0)|0\ra e^{ip\cdot x}
\end{aligned}
\end{equation*}
Inverting the Fourier transform yields
\begin{equation}
  \rho(p^2)\la 0|\mathcal{O}^\dagger_{\mu\nu}(0)|p\ra\la p|O_{\rho\sigma}(0)|0\ra=\int d^4x e^{-ip\cdot x} \la\mathcal{O}^\dagger_{\mu\nu}(x)\mathcal{O}_{\rho\sigma}(0)\ra
\end{equation}
The (Euclidean) position-space two point functions in a CFT is fully fixed up to an overall normalization using conformal invariance and dimensional analysis. For the antisymmetric 2-tensor, it is given by \cite{Grinstein:2008qk}
\begin{equation}
\begin{aligned}
  \la\mathcal{O}^\dagger_{\mu\nu}(x)\mathcal{O}_{\rho\sigma}(0)\ra=&C_{AT}\frac{1}{(2\pi)^2}\frac{(I_{\mu\rho}(x)I_{\nu\sigma}(x)-\frac{1}{4}g_{\mu\nu}g_{\rho\sigma})-(\mu\leftrightarrow\nu)}{(x^2)^{d}},
\end{aligned}
\end{equation}
where $d$ is the scaling dimension of the operator $\mathcal{O}_{\mu\nu}$, $C_{AT}$ is an overall normalization, and
\begin{equation*}
  I_{\mu\nu}=g_{\mu\nu}-2\frac{x_\mu x_\nu}{x^2}.
\end{equation*}
The Fourier transform is given by 
\begin{equation}
  \begin{aligned}
    \la\mathcal{O}^\dagger_{\mu\nu}(x)\mathcal{O}_{\rho\sigma}(0)\ra=&\int \frac{d^4k}{(2\pi)^4}e^{ik\cdot x}\Bigg[C_{AT}(-1)\frac{\Gamma(3-d)}{4^{d-1}\Gamma(d+1)}(k^2)^{d-2} \\ & \hspace{-1cm} \times\Bigg(\big(g_{\mu\rho}g_{\nu\sigma}-(\mu\leftrightarrow\nu)\big)
    -2\Big(g_{\mu\rho}\frac{k_\nu k_\sigma}{k^2}+g_{\nu\sigma}\frac{k_\mu k_\rho}{k^2}-(\mu\leftrightarrow\nu)\Big)\Bigg)\Bigg]
  \end{aligned}
\end{equation}
We analytically continue the above Euclidean expressions to Minkowski space in the mostly minus signature.
\begin{equation}
\begin{aligned}
  \rho(p^2)\la 0|\mathcal{O}^\dagger_{\mu\nu}(0)|p\ra\la p|O_{\rho\sigma}(0)|0\ra=&C_{AT}(-1)\frac{\Gamma(3-d)}{4^{d-1}\Gamma(d+1)}(-p^2)^{d-2}\\
  &\hspace{-3cm}\times\Bigg(\big(g_{\mu\rho}g_{\nu\sigma}-(\mu\leftrightarrow\nu)\big)-2\Big(g_{\mu\rho}\frac{p_\nu p_\sigma}{p^2}+g_{\nu\sigma}\frac{p_\mu p_\rho}{p^2}-(\mu\leftrightarrow\nu)\Big)\Bigg)
\end{aligned}
\end{equation}
To have an unparticle interpretation for the state generated by $\mathcal{O}_{\mu\nu}$, we need to choose the normalization such that the prefactor of $p^2$ corresponds to the phase space of $d$ massless particles, i.e. choose $C_{AT}$ such that the following relation holds:
\begin{equation}
  (-1)^{d-2}C_{AT}\frac{\Gamma(3-d)}{4^{d-1}\Gamma(d+1)}\equiv A_d=\frac{16\pi^{5/2}}{(2\pi)^{2d}}\frac{\Gamma(d+1/2)}{\Gamma(d-1)\Gamma(2d)}. 
\label{eq:A_d}
\end{equation}
We now perform the index contractions.
\begin{equation*}
\begin{aligned}
  \rho(p^2)\sum |\mathcal{M}|^2&\propto p_{1\beta}p_{2\gamma}p^\rho p^\alpha\left[\delta^\beta_\rho\delta^\gamma_\alpha+\delta^\beta_\alpha\delta^\gamma_\rho-g^{\beta\gamma}g_{\alpha\rho}-\frac{4}{p^2}(g_{\alpha\rho}p^\beta p^\gamma+g^{\beta\gamma}p_\alpha p_\rho)\right]\\
  &=-\left(2(p_1\cdot p)(p_2\cdot p)+5(p_1\cdot p_2)p^2\right).
\end{aligned}
\end{equation*}
Using the delta function and the fact that the particles are massless, we have the following:
\begin{equation*}
  (p_1+p_2)^2=p^2=2p_1\cdot p_2,\qquad(p-p_2)^2=p_1^2=0=p^2-2p\cdot p_2,~
\end{equation*}
\begin{equation*}
  (p-p_1)^2=p_2^2=0=p^2-2p\cdot p_1
\end{equation*}
Thus,
\begin{equation}
  \rho(p^2)\sum|\mathcal{M}|^2=A_d \frac{12 e^2\lambda^2}{\Lambda^{2d-4}}(p^2)^{d-2}
\end{equation}
Assuming the SM particles follow the Maxwell-Boltzmann distribution, we get
\begin{equation}
\begin{aligned}
  n_1 n_2\la \sigma v E\ra=&A_d\frac{48e^2\lambda^2}{\Lambda^{2d-4}}\frac{1}{(2\pi)^6}\int d^4p (p^2)^{d-2}\theta(p^2)\theta(p^0)p^0 e^{-p_0/T}\\
  &\times\int d^4 p_1 d^4p_2 \delta(p_1^2)\delta(p_2^2)\theta(p_1^0)\theta(p_2^0)\delta^4(p_1+p_2-p)
\end{aligned}
\end{equation}
Computing the remaining integral yields
\begin{equation}
  n_1 n_2\la\sigma v E\ra=A_d\frac{48 e^2\lambda^2}{4(2\pi)^5\Lambda^{2d-4}}\frac{1}{2}\frac{\Gamma(\frac{3}{2})\Gamma(d-1)}{\Gamma(d+\frac{1}{2})}\Gamma(2d+1)T^{2d+1}
\end{equation}

\subsection{Higgs annihilation}
\label{subapp:HH_ann}

Next, we consider freeze-in through $H^\dagger H\rightarrow \text{CFT}$ (assuming only kinetic mixing through $U(1)_Y$ between the SM and the CFT sector). The matrix element is
\begin{equation*}
  \mathcal{M}=-\frac{2\lambda g}{\Lambda^{d-2}}\frac{1}{p^2}(p_1-p_2)_\mu p_\nu \la p|\mathcal{O}^{\mu\nu}|0\ra
\end{equation*}

Repeating the above in the massless limit yields

\begin{equation}
  \rho(p^2)\sum |\mathcal{M}|^2=\frac{4\lambda^2g^2}{\Lambda^{2d-4}}A_d (p^2)^{d-2}
\end{equation}

After EWSB, the above process is matched onto $Z h\rightarrow\text{CFT}$ which then gets quickly shut off once the Higgs boson decouples from the thermal bath.

\subsection{Gauge boson initial state}
Due to thermal effects, the longitudinal mode of the photon picks up a thermal mass proportional to $T$. This allows the ``decay'' process of gauge bosons into unparticles. The matrix element for $A^\mu\rightarrow \text{CFT}$ is given by
\begin{equation}
  \mathcal{M}=-\frac{2\lambda}{\Lambda^{d-2}}\epsilon^\mu p^\nu \la p|\mathcal{O}_{\mu\nu}|0\ra
\end{equation}
Squaring and performing the polarization sum yields
\begin{equation}
  \sum |\mathcal{M}|^2=\frac{4\lambda^2}{\Lambda^{2d-4}}\left(-g^{\mu\sigma}+\cancel{\frac{p^\mu p^\sigma}{p^2}}\right)p^\nu p^\rho\la 0|\mathcal{O}^\dagger_{\rho\sigma}|p\ra\la p|\mathcal{O}_{\mu\nu}|0\ra
\end{equation}
where the second term vanishes by the antisymmetric property of $\mathcal{O}_{\mu\nu}$\footnote{Here, we used the polarization vectors for a massive gauge field. They are different from the dressed polarization vectors for photons in a thermal bath. This will generically result in functions of $p$ arising in front of $g^{\mu\sigma}$ and $p^\mu p^\sigma$.  This does not affect the fact that the second term cancels.}. The rate of energy density transfer via this decay process is given by
\begin{equation}
  \left(\int d\Pi_A f({p_A})\right)\int \frac{d^4 p}{(2\pi)^4}\rho(p^2)(2\pi)^4\delta^4(p_A-p)
  \left(\sum|\mathcal{M}|^2\right) (E_A)
\end{equation}
As was done in the case of the fermion pair annihilation, we replace the ``momentum-space wavefunctions'' with the two-point function and use the correct normalization to yield the unparticle interpretation. So the right-most integral is equal to
\begin{equation}
\begin{aligned}
  \rho(p^2)\left(\sum|\mathcal{M}|^2\right)=& \frac{4 \lambda^2}{\Lambda^{2d-4}}A_d (p^2)^{d-2}g^{\mu\sigma}p^\nu p^\rho\\
 & \hspace{-1cm} \times\left((g_{\mu\rho}g_{\sigma\nu}-(\mu\leftrightarrow\nu))-2\left(g_{\mu\rho}\frac{p_\mu p_\sigma}{p^2}+g_{\nu\sigma}\frac{p_\mu p_\rho}{p^2}-(\mu\leftrightarrow\nu)\right)\right)
\end{aligned}
\end{equation}
Performing the index contractions yield
\begin{equation}
  \rho(p^2)\left(\sum|\mathcal{M}|^2\right)= \frac{4\lambda^2}{\Lambda^{2d-4}}A_d (p^2)^{d-2}(3p^2)
\end{equation}
Thus, the collision term is
\begin{equation}
  \left(\int d\Pi f(p)\right)\frac{4 p^0 \lambda^2}{\Lambda^{2d-4}}A_d (p^2)^{d-2}(3p^2)
\end{equation}
The phase space integral includes a delta function which enforces the on-shell relation. Using that, the energy transfer rate simplifies down to
\begin{equation}
  \frac{6\lambda^2 A_d m^{2d-2}}{\Lambda^{2d-4}}\int \frac{d^3 p}{(2\pi)^3} f(p)
\end{equation}
The final quantity is precisely the number density of a Boltzmann-distributed particle at temperature $T$. This is given by
\begin{equation}
  \int \frac{d^3 p}{(2\pi)^3} f(p)=\frac{1}{\pi^2}T^3
\end{equation}
Thus,
\begin{equation}
  \Gamma = \frac{6\lambda^2A_d}{\Lambda^{2d-4}}m(T)^{2d-2} \frac{1}{\pi^2}T^3
\end{equation}

\section{Details of Stellar Cooling Estimates}
\label{app:star_cooling}

Here, we will discuss the estimation of the energy density loss rate in most stellar systems.
A more detailed discussion of stellar evolution bound on COFI theories can be found in \cite{Hong:2022gzo}. 

\subsection{Main sequence and horizontal branch stars}
In both main sequence and horizontal branch stars, the dominant mechanism for energy loss is via an analog of the Compton process \cite{Raffelt:1996wa}. An incoming photon is absorbed by an electron which subsequently radiates either DM pairs or unparticles.

In order for DM pairs to be directly produced, the temperature (or equivalently the scale of momentum transfer) must be below $m_\text{gap}$. We can integrate out the dark photon to obtain the effective operator
\begin{equation}
  \mathcal{L}\supset i\frac{\epsilon e c_w g_s}{m_\text{gap}^2}(\bar{e}\gamma^\mu e)\left(\pi^+\overset{\leftrightarrow}{\partial}_\mu\pi^-\right)\equiv i\frac{4G_\text{eff}}{\sqrt{2}}(\bar{e}\gamma^\mu e)\left(\pi^+\overset{\leftrightarrow}{\partial}_\mu\pi^-\right)
\end{equation}
Up to some $\mathcal{O}(1)$ factors from the difference in particle statistics, the kinematic factor of the spin-averaged amplitude is the same as that from neutrino pair emission. As such, one can take the existing computation of the energy density loss rate from neutrino emission and perform the replacement $G_F\rightarrow G_\text{eff}$. This gives
\begin{equation}
  \dot{\varepsilon}_\text{Comp, h}\simeq\frac{7!}{\pi^2}\left(\frac{Y_e}{m_N}\right)\frac{\alpha}{8\pi^2}G_\text{eff}^2\frac{T^8}{m_e^2},\label{eq:comp}
\end{equation}
where $Y_e$ is the electron to nucleon ratio and $m_N$ is the nucleon mass.

For unparticle production which occurs when $T>m_\text{gap}$, while the estimate is less robust, nevertheless a reasonable estimation is possible. This is done by rescaling the result of the energy loss via emission of a light scalar from the same Compton-like process \cite{Hong:2022gzo}. The main difference between the two processes are the number of final state particles, average energy carried away by the final states and the couplings. Noting this, we can write
\begin{equation}
  \frac{\dot{\varepsilon}_\text{Comp, CFT}}{\dot{\varepsilon}_\text{scalar}}\sim\frac{(e\lambda/\Lambda^{d-2})^2}{g^2}\frac{A_{d+1}}{A_{2}}T^{2d-4},
\end{equation}
where $g$ is the Yukawa coupling of the electrons with the light scalar. The factors of $T$ were added to ensure that the RHS is dimensionless.\footnote{Here, $T$ is the right dimensionful parameter to balance the dimensions since the characteristic energy transfer is controlled by $T$. This is not always true. For example, in the case of electron-positron annihilation at the core of supernova, one has to use the fermi energy $E_F$ instead. For more details, see \cite{Hong:2022gzo}.} Plugging in the known result for $\dot{\varepsilon}_\text{scalar}$ \cite{Raffelt:1996wa} gives 
\begin{equation}
  {\dot{\varepsilon}_\text{Comp, CFT}}\sim\alpha^2\lambda^2 A_{d+1}\frac{T^{2d}}{\Lambda^{2d-4}}\frac{Y_e}{m_u m_e^2}.
\end{equation}
Here, we are missing potentially important numerical factors which may affect the bounds, but this is beyond the scope of this paper.

\subsection{SN 1987A}

In supernova progenitor cores, the dominant energy loss mechanism is nuclear brems-\\strahlung \cite{Chang:2018rso}. Since the nucleons are nearly degenerate in the core, the typical energy scale is $p_F\approx\sqrt{m_NT}$. Using a similar process as before, we can estimate the rate of energy loss by producing CFT states by rescaling the rate via emission of a light scalar
\begin{equation}
  \frac{\dot{\varepsilon}_\text{brem, CFT}}{\dot{\varepsilon}_\text{scalar}}\sim \frac{(e\lambda/\Lambda^{d-2})^2}{g^2}\frac{A_{d+2}}{A_3}p_F^{2d-4},
\end{equation}
where $g$ is now the Yukawa coupling of a nucleon pair to the light scalar. Plugging in the known result for $\dot{\varepsilon}_\text{scalar}$ \cite{Raffelt:1996wa}, we get
\begin{equation}
  \dot{\varepsilon}_\text{brem, CFT}\sim \frac{(e\lambda/\Lambda^{d-2})^2}{4\pi}\alpha_\pi^2\frac{44}{15^3}\frac{A_{d+2}}{A_3}\left(\frac{T}{m_N}\right)^4 p_F^5 G_\text{scalar}(m_\pi/p_F),
\end{equation}
where $\alpha_\pi\approx 15$ is the coupling of nucleons to pions and $G_\text{scalar}$ is the correction to the bremsstrahlung rate for nonzero pion mass. For the density of the progenitor star core, $G_\text{scalar}(m_\pi/p_F)\approx0.8$.

\bibliographystyle{JHEP}
\bibliography{refs}
\end{document}